\documentclass[12pt]{iopart}

\usepackage{iopams}
\usepackage{graphicx}
\usepackage{xcolor}
\usepackage{ulem}
\usepackage{hyperref}  
\usepackage{url}       

\newcommand{\lequivt}{\mathcal{L}_{\mathrm{equiv}}}
\newcommand{\lequivs}{l_{\mathrm{equiv}}}
\newcommand{\lid}{l_{\hat{I}}}
\newcommand{\lrep}{\mathcal{L}_{\mathrm{rep}}}
\newcommand{\ntr}{N_{\mathrm{T}}} 
\newcommand{\nst}{N_{\mathrm{S}}}
\newcommand{\ynn}{\vec{y}_{\mathrm{NN}}}

\begin{document}

\title{Learning finite symmetry groups of dynamical systems via equivariance detection}

\author{Pablo Calvo-Barlés$^{1,2}$, Sergio G. Rodrigo$^{1,3}$ and Luis Martín-Moreno$^{1,2}$}

\address{$^1$Instituto de Nanociencia y Materiales de Aragón (INMA), CSIC-Universidad de Zaragoza, Zaragoza 50009, Spain}
\address{$^2$Departamento de Física de la Materia Condensada, Universidad de Zaragoza, Zaragoza 50009, Spain}
\address{$^3$Departamento de Física Aplicada, Universidad de Zaragoza, Zaragoza 50009, Spain}
\ead{lmm@unizar.es}
\vspace{10pt}
\begin{indented}
\item[]November 2024
\end{indented}

\begin{abstract}
  In this work, we introduce the Equivariance Seeker Model (ESM), a data-driven method for discovering the underlying finite equivariant symmetry group of an arbitrary function. ESM achieves this by optimizing a loss function that balances equivariance preservation with the penalization of redundant solutions, ensuring the complete and accurate identification of all symmetry transformations. We apply this framework specifically to dynamical systems, identifying their symmetry groups directly from observed trajectory data. To demonstrate its versatility, we test ESM on multiple systems in two distinct scenarios: (i) when the governing equations are known theoretically and (ii) when they are unknown, and the equivariance finding relies solely on observed data. The latter case highlights ESM’s fully data-driven capability, as it requires no prior knowledge of the system’s equations to operate.
\end{abstract}

%
%
%
%
%


\section{Introduction}\label{sec:intro}

Symmetries are fundamental to the physical sciences, serving as powerful tools to reduce complexity and reveal the underlying laws governing physical phenomena \cite{gross1996therole, noether1918invariante}. Discovering the underlying symmetries of a physical system when they are not explicitly known is both highly valuable and very challenging, especially when only observational data is available. Machine learning (ML), and particularly neural networks (NNs), provides a framework for automatic symmetry discovery due to its capacity to identify patterns from data. Recent advancements have shown that ML models can uncover physical insights from data in unknown or partially known physical systems \cite{wang2023scientific,yu2024learning}, including governing equations of dynamical systems \cite{brunton2016discovering, rudy2017datadriven}, conservation laws \cite{liu2021machine,liu2022machine,liu2024interpretable,ha2021discovering,wetzel2020discovering,lu2023discovering}, or physically relevant quantitites \cite{iten2020discovering,greydanus2019hamiltonian,cranmer2020lagrangian}. Various aspects of ML-assisted symmetry discovery have been explored using different strategies. For instance, NNs have been used for finding coordinate transformations that reveal hidden symmetries \cite{liu2022machinelearninghidden,bondesan2019learning}, learning transformations that preserve the statistical distribution of data sets \cite{desai2022symmetry, yang2023generative}, recognizing symmetries from embedding layers \cite{barenboim2021symmetry,krippendorf2020detecting} and identifying pairs of symmetric events \cite{decelle2019learning}.


Much of the research in symmetry discovery has concentrated on continuous symmetry groups, particularly the learning of generators associated with Lie groups \cite{gabel2023learning,gabel2024datadriven,ko2024learning,forestano2023deep,dehmamy2021automatic,yang2023generative,hou2024machine}. However, there has been comparatively less focus on discovering finite symmetry groups. Some methods for continuous group detection have demonstrated the capability to identify individual transformations of finite groups \cite{yang2023generative}, but they do not guarantee the identification of all the elements within the group. In contrast, the discovery of the full finite symmetry group has been addressed in \cite{calvo2024finding}, but the method is limited to invariance detection, that is, to identify transformations in mappings $\vec{y} (\vec{x})$ that act on the input $\vec{x}$ and leave the output $\vec{y}$ unchanged. The simplest example of an invariant function is that of an even function, $y(-x) = y(x)$. But there are symmetric functions that are not invariant, as for instance an odd function $y(-x) = -y(x)$. These are called equivariant functions, meaning that applying a transformation to the input induces a corresponding and predictable transformation of the output \cite{bronstein2017geometric,cohen2016group,satorras2021enequivariant,cohen2019gauge,thomas2018tensor}.


In this article, we extend the work on equivariance detection by focusing on symmetries inherent in the equations governing dynamical systems. In this context, these symmetries manifest as a specific type of equivariance in which the transformation applied to the input is identical to that applied to the output. Symmetries are particularly relevant in dynamical systems. For example, symmetry considerations can be used to simplify the spectral structure of the Koopman operator—an infinite-dimensional linear operator that evolves measurement functions of state space variables—and calculate more efficiently its eigenmodes \cite{salova2019koopman}. We propose a method to automatically detect the full finite group of linear symmetries in arbitrary nonlinear dynamical systems from collected trajectory data.


The text is structured as follows: in section \ref{sec:math_def}, we state the mathematical definition of the problem we are interested in. Next, in section \ref{sec:ESM_method}, we introduce the general ML framework for identifying equivariant symmetry groups from data. In section \ref{sec:ESM_proc}, we illustrate our method with a one-dimensional simple example. In section \ref{sec:app_compl}, we demonstrate our method's effectiveness on well-known dynamical systems. Finally, we conclude in section \ref{sec:conc} with a summary of our findings.


\section{Mathematical definition of the problem}\label{sec:math_def}


We consider nonlinear dynamical systems described by ordinary differential equations (ODEs):
\begin{equation}\label{eq:dyn_sys}
  \frac{d \vec{x}}{dt} = \vec{y}(\vec{x}),
\end{equation}
where $\vec{x} (t) \in \mathbb{R}^n$ denotes the system's state evolving over time $t$. The time derivative of the state is provided by the nonlinear vector field $\vec{y}\left( \vec{x} (t)\right): \mathbb{R}^n \rightarrow \mathbb{R}^n$, that thus determines the system dynamical evolution.


A linear symmetry of the dynamical system in Eq. \ref{eq:dyn_sys} is a matrix $\hat{D} \in \mathbb{R}^{n \times n}$ such that if $\vec{x}(t)$ is a solution, then $\hat{D} \vec{x}(t)$ is also a solution. For the system to exhibit this symmetry, the function $\vec{y}$ must be equivariant under the action of $\hat{D}$, meaning:
\begin{equation}\label{eq:sym_def}
\vec{y}(\vec{x}) = \hat{D}^{-1} \vec{y}( \hat{D} \vec{x} ), \; \forall \vec{x} \in \mathbb{R}^n.
\end{equation}


We consider dynamical systems presenting a finite symmetry group of order $K$. This means that there exists a unique set of matrices $\{ \hat{D}_{\alpha} \}$, for $\alpha = 1, \dots, K$, such that each matrix satisfies the equivariance condition in Eq. \ref{eq:sym_def}.


The goal of this study is to identify all $K$ elements of the symmetry group from a set of $\ntr$ trajectories of states $\{ \vec{x}^{(r)}(t_s) \}$ and time derivatives $\{ \vec{y}^{(r)}(t_s) \}$ ($r = 1, \dots, \ntr$). Each trajectory correspond to a different initial condition and is observed at $\nst$ equidistant time steps $t_s = s\Delta t$, with $s = 1, \dots, \nst$ and $\Delta t$ being the time interval between observations.


The time derivatives $\{ \vec{y}^{(r)}(t_s) \}$ can be obtained either analytically (if $\vec{y}(\vec{x})$ is known), measured, or numerically approximated. In this study, we consider two hypothetical scenarios: (1) when the analytical expression of $\vec{y}(\vec{x})$ is known, allowing direct computation of $\vec{y}^{(r)}(t_s) \equiv \vec{y}[\vec{x}^{(r)}(t_s)]$, and (2) a data-driven case where $\vec{y}(\vec{x})$ is unknown, so the derivatives are approximated as $\vec{y}^{(r)}(t_s) \equiv \left[\vec{x}^{(r)}(t_{s+1}) - \vec{x}^{(r)}(t_s)\right]/\Delta t$. This last scenario will be used to demonstrate that the method can operate purely in a data-driven manner, without requiring the governing equations.


\section{A method for detecting equivariant symmetry groups}\label{sec:ESM_method}

In this section we explain the general framework of the technique proposed in this paper to find the symmetry group of a given dynamical system. This includes the ML model architecture, the loss functions and the final refinement used to obtain the results.


\subsection{Equivariance Seeker Model architecture}\label{sec:ESM_arch}

\begin{figure*}
  \centering
  \includegraphics[width=\textwidth]{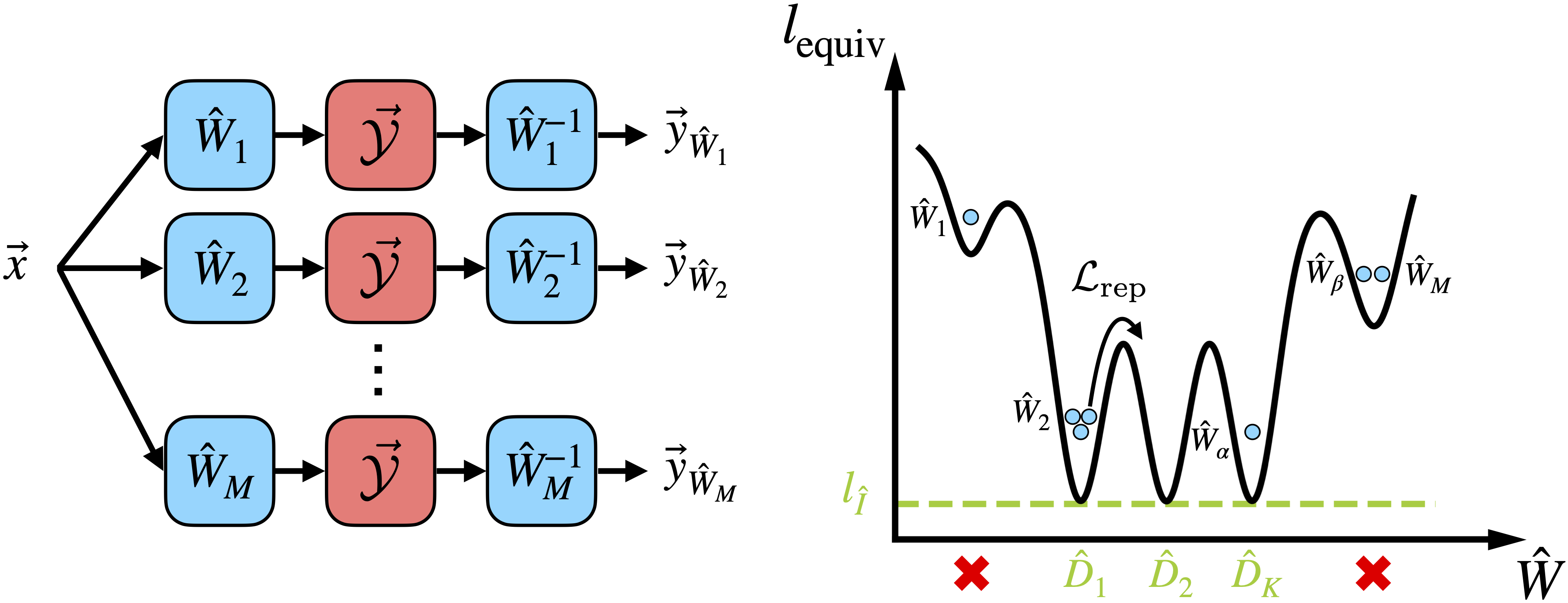}
  \caption{Schematic representation of the ESM and its loss function. Left: illustration of ESM's processing flow: the input state $\vec{x}$ is operated by $M$ parallel branches. Each $\alpha$ branch applies three consecutive blocks: first, a trainable matrix $\hat{W}_{\alpha}$. Second, the non-trainable function $\vec{\mathcal{Y}}$ that produces the dynamical system's time derivative. Third, the matrix $\hat{W}_{\alpha}^{-1}$. The ESM outputs $M$ predictions $\vec{y}_{\hat{W}_{\alpha}} = \hat{W}_{\alpha}^{-1} \vec{\mathcal{Y}} \left( \hat{W}_{\alpha} \vec{x} \right)$. Right: schematic illustration of the single-branch equivariance loss function. It has $K$ global minima of the same magnitude (zero or $\lid$), corresponding to the symmetry transformations $\hat{D}_{\alpha}$. The function may present other local minima of higher magnitude, which are not associated to symmetries. Branch matrices $\hat{W}_{\alpha}$ (for $\alpha = 1, \dots, M$) are independently optimized with the equivariance loss and, at the same time, jointly optimized with the repetition loss.}
  \label{fig:FIG_1}
\end{figure*}


We introduce the Equivariance Seeker Model (ESM) [depicted in Fig.~\ref{fig:FIG_1}(a)], which processes an input state $\vec{x}$ through $M$ parallel branches, indexed by $\alpha = 1, \dots , M$ ($M$ is an hyperparameter). Each branch consists of three consecutive blocks: (1) a trainable matrix $\hat{W}_{\alpha} \in \mathbb{R}^{n \times n}$, (2) a non-trainable block $\vec{\mathcal{Y}}$, identical for all branches, which replicates the system's derivative vector field (Eq.~\ref{eq:dyn_sys}), and (3) the inverse matrix of the first block, $\hat{W}_{\alpha}^{-1}$. After processing, the ESM outputs $M$ predictions, denoted as
\begin{equation}\label{eq:outputs}
  \vec{y}_{\hat{W}_{\alpha}} (\vec{x}) = \hat{W}_{\alpha}^{-1} \vec{\mathcal{Y}} \left( \hat{W}_{\alpha} \vec{x} \right),
\end{equation}
for $\alpha = 1, \dots, M$. Note that the trainable weights of the ESM correspond to the set of branch matrices $\{ \hat{W}_{\alpha} \}$.


The derivative vector field block $\vec{\mathcal{Y}}$ is defined based on prior knowledge of the system. If the analytical differential equation Eq.~\ref{eq:dyn_sys} is known, $\vec{\mathcal{Y}} (\vec{x}) \equiv \vec{y}(\vec{x})$. Otherwise, an oracle model is pre-trained on data pairs $\{ \vec{x}^{(r)} (t_s) , \vec{y}^{(r)} (t_s) \}$ to predict the derivative from an arbitrarily chosen state, including those not within the training data. In our implementation, we use as oracle a simple feed-forward NN, denoted as $\ynn$, such that $\vec{\mathcal{Y}} (\vec{x}) \equiv \ynn (\vec{x})$. However, the framework is flexible and can incorporate other ML-based oracles \cite{brunton2016discovering,schmid2010dynamic,chen2018neural}.


\subsection{Equivariance loss function}\label{sec:equiv_loss}


To identify the ground-truth symmetry transformations $\{ \hat{D}_{\alpha} \}$, we optimize the weights $\{ \hat{W}_{\alpha} \}$ such that each branch approximates the vector field $\vec{y}(\vec{x})$. When this goal is achieved, the converged branch matrices represent symmetry transformations, as they satisfy the equivariance condition given by Eq.~\ref{eq:sym_def}.

We define the equivariance loss as
\begin{equation}\label{eq:l_equiv}
  \mathcal{L}_{\mathrm{equiv}} \left( \{ \hat{W}_{\alpha} \} \right) = \frac{1}{M} \sum_{\alpha = 1}^{M} l_{\mathrm{equiv}} ( \hat{W}_{ \alpha } ),
\end{equation}
which depends on the set of all branch matrices and is the average of the single-branch equivariance losses, given by
\begin{equation}\label{eq:l_equiv_1}
  l_{\mathrm{equiv}} \left( \hat{W} \right) = \frac{1}{N} \sum_{i=1}^{N} || \vec{y}_i - \vec{y}_{\hat{W}} ( \vec{x}_i  ) ||^2,
\end{equation}
where $|| \cdot ||$ denotes the $L_2$ norm. The dataset $\{ \vec{x}^{(r)}(t_s), \vec{y}^{(r)}(t_s) \}$ is randomly shuffled and split into three parts: two subsets are used for the oracle training and validation, while the third subset is used for ESM training. In the formula above, $N$ represents the number of samples assigned to ESM training and $i$ is an index that combines trajectories and time steps for simplicity.


The function $\lequivs$ [sketched in Fig. \ref{fig:FIG_1}(b)] depends on a single matrix $\hat{W}$, is non-negative and reaches its minimum value when $\hat{W}$ is a symmetry transformation. Consequently, $\lequivs$ has $K$ global minima, each corresponding to one of the ground-truth symmetry transformations $\hat{D}_{\alpha}$. Indeed, if the analytical derivative is available [$\vec{\mathcal{Y}} (\vec{x}) \equiv \vec{y}(\vec{x})$], the value of $\lequivs$ in these minima is zero because of Eq.~\ref{eq:sym_def}.  In the absence of this analytical expression  [$\vec{\mathcal{Y}} (\vec{x}) \equiv \ynn(\vec{x})$], the values of $\lequivs$ in global minima become of the order of the oracle's prediction error (which is assumed to be small). Still, we expect all $K$ global minima to have approximately the same value, which can be estimated by evaluating $\lequivs$ at the identity transformation $\hat{I}$, i.e., the trivial symmetry. This value at the identity, that we denote as $\lid$, coincides with the oracle's mean squared error (MSE),
\begin{equation}\label{eq:oracle_mse}
  \lid = \left. l_{\mathrm{equiv}} \left( \hat{I} \right) \right|_{\vec{\mathcal{Y}} = \ynn} = \frac{1}{N} \sum_{i=1}^{N} || \vec{y}_i - \ynn ( \vec{x}_i  ) ||^2.
\end{equation}


\subsection{Repetition loss function}\label{sec:rep_loss}

The equivariance loss in Eq.~\ref{eq:l_equiv} reaches its minimum when each ESM branch identifies a symmetry transformation, meaning that each $\hat{W}_{\alpha}$ converges to one of the $K$ global minima of $\lequivs$. However, $\lequivt$ does not enforce the discovery of all distinct symmetry transformations within the group. For instance, nothing in $\lequivt$ prevents all branch matrices converging to the same symmetry transformation (e.g., $\hat{W}_{\alpha} = \hat{I} \; \forall \alpha$) leaving the remaining $K-1$ minima unexplored.


We propose a training methodology for the ESM to overcome this limitation and discover the entire finite symmetry group (all $K$ elements) in a single run. To ensure all global minima of $\lequivs$ are found, we introduce the so-called repetition loss term (defined below), that penalizes branch matrix repetition during training. First, we assume that the number of branches satisfies $M \geq K$. At the end of the process we will be able to check that the symmetry group has been found, so there is always the possibility to repeat the process with a larger $M$, if initially the chosen $M$ is not large enough. Conversely, if $M$ is large enough, pruning process removes redundant branches during training, stopping at $M = K$, as will be described later in subsection \ref{sec:local_redundant}.


We introduce the repetition loss term, defined as
\begin{equation}\label{eq:l_rep}
  \mathcal{L}_{\mathrm{rep}} \left( \{ \hat{W}_{\alpha} \} \right) = A \sum_{\alpha = 1}^{M} \sum_{\beta = 1}^{\alpha - 1} \exp{ \left( -\frac{1}{\sigma} || \hat{W}_{\alpha} - \hat{W}_{\beta} ||_{F}^2 \right) },
\end{equation}
that depends on two hyperparameters: amplitude $A$ and width $\sigma$. Here, $|| \cdot ||_{F}$ denotes the Frobenius norm \cite{brunton2019datadriven}, which can be interpreted as a distance in the matrix space $\mathbb{R}^{n \times n}$. The repetition loss penalizes similarity between branch matrices. It consists of a sum over all distinct indices $\alpha \neq \beta$, with each term being a Gaussian function of the distance between $\hat{W}_{\alpha}$ and $\hat{W}_{\beta}$. The penalty is maximum when two matrices are identical ($\hat{W}_{\alpha} = \hat{W}_{\beta}$) and decreases as their distance increases. Intuitively, if two branch matrices are attracted to the same minimum of $\lequivs$, the repetition loss generates a "repulsive action" between them, with its strength controlled by $A$ and the action range determined by $\sigma$.


The total loss function combines equivariance and repetition losses,
\begin{equation}\label{eq:loss}
  \mathcal{L} = \mathcal{L}_{\mathrm{equiv}} + \mathcal{L}_{\mathrm{rep}}.
\end{equation}
This dual approach promotes symmetry detection in such a way that branch matrices are forced to converge at distinct global minima.


Nevertheless, a naive ESM training with $\mathcal{L}$ using fixed $\sigma$ and $A$ hyperparameters is not a good solution. The influence of $\lrep$ (controlled by $\sigma$ and $A$) is highly sensitive to the $\lequivs$ landscape, which depends on the dynamical system and collected data. A small $\sigma$ limits the range of $\lrep$, allowing multiple branches to fall into the basin of the same minimum, thus failing to separate branch matrices. Conversely, a large $\sigma$ produces repulsion even between matrices located in different global minima, moving them away from the minimum of $\lequivt$, thus yielding to inaccurate symmetry prediction.


To overcome the pitfalls of a naive training with $\mathcal{L}$, we follow a protocole where the hyperparameters $\sigma$ and $A$ are modified throughout training according to a predefined function of the epochs, gradually diminishing the influence of $\lrep$ until it vanishes. By the end of training, only the equivariance loss remains ($\mathcal{L} \approx \lequivt$), ensuring the optimization focuses solely on symmetry detection. This protocole will be illustrated in section \ref{sec:ESM_method} with a simple example.


\subsection{Branch-removal processes: local minima and redundant solutions}\label{sec:local_redundant}

As previously mentioned, we expect that the function $\lequivs$ presents global minima having—exactly or approximately—the same value for all matrices that are exact symmetries of the dynamical system. Additionally, $\lequivs$ may also exhibit local minima for matrices that do not represent symmetries. These local minima appear for loss values significantly larger than those of the global minima. The optimization process can occasionally cause the branch matrices to become trapped in these local minima, creating an obstacle for identifying true symmetries.

To mitigate this, our method includes a process at the last stages of training (after $\lrep$ vanishes) in which we remove branches that have converged to local minima. For that, we calculate each single-branch equivariance loss, $\lequivs (\hat{W}_{\alpha})$, and eliminate those branches $\alpha$ whose values are significantly higher than the function's lower bound. We thus use tolerance $\varepsilon_{\mathrm{equiv}}$ so that, if $\lequivs (\hat{W}_{\alpha}) > \varepsilon_{\mathrm{equiv}}$, the branch $\alpha$ is removed. The value of this tolerance depends on the available knowledge of the system:

\begin{enumerate}
  \item In the case of $\vec{\mathcal{Y}} (\vec{x}) = \vec{y}(\vec{x})$, since the lower bound of $\lequivs$ is zero, we aim to discard any branch matrix $\hat{W}_{\alpha}$ that does not satisfy $\lequivs (\hat{W}_{\alpha}) = 0$. However, converged branch matrices do not strictly reach this value in practice due to gradient descent oscillations around the minimum. For this reason, we use a tolerance $\varepsilon_{\mathrm{equiv}} = 10^{-6}$, which is several orders of magnitude smaller than typical values of $\lequivs$ in our numerical experiments.
  
  \item In the case of $\vec{\mathcal{Y}} (\vec{x}) = \ynn (\vec{x})$, the value $\lequivs (\hat{D}_{\alpha})$ of each global minimum is equal to reference value $\lid$ given by the identity matrix, up to a fluctuation. This fluctuation arises because calculating $\lequivs (\hat{D}_{\alpha})$ requires the interpolation of $\ynn$ at states $\hat{D}_{\alpha}\vec{x}_i$, which do not belong to the original ESM training set. Therefore, the quantity $|| \vec{y}_i - \hat{D}_{\alpha}^{-1} \ynn (\hat{D}_{\alpha}\vec{x}_i)||^2$ differs from the oracle's error $|| \vec{y}_i - \ynn (\vec{x}_i)||^2$ for each sample $i$. These individual differences result in $\lequivs (\hat{D}_{\alpha}) \neq \lid$. Nevertheless, if the oracle has sufficiently low generalization error, the values $|| \vec{y}_i - \hat{D}_{\alpha}^{-1} \ynn (\hat{D}_{\alpha}\vec{x}_i)||^2$ [and therefore $\lequivs (\hat{D}_{\alpha})$] should lie within the oracle's error distribution. For this reason, we use a tolerance given by the maximum error among the ESM training set, i.e., $\varepsilon_{\mathrm{equiv}} = \max_{i = 1}^N  \{ || \vec{y}_i - \ynn (\vec{x}_i)||^2$ \}.
\end{enumerate}

Once the local minima branches are pruned, all ESM branch matrices should represent a given symmetry transformation. These are maximally separated thanks to the previous $\lrep$ action, promoting the identification of all possible symmetries of the group. As mentioned above, the number of branches should initially exceed the order of the group $K$, which means that some symmetries $\hat{D}_{\alpha}$ can inevitably be found more than once. We thus perform a second final process in which redundant branches are removed. To do so, we calculate the quantities $d (\hat{W}_{\alpha}, \hat{W}_{\beta})$ for all pairs of branches $\alpha, \beta$, with the average element-wise distance (AED) between two matrices $\hat{A}$ and $\hat{B}$ defined as
\begin{equation}\label{eq:matrix_d}
  d \left( \hat{A} , \hat{B} \right) = \frac{1}{n^2} \sum_{k l} |\hat{A}_{kl} - \hat{B}_{kl}|,
\end{equation}
where $n$ is the matrix dimension, and $k,l$ index the matrix elements. Unlike the Frobenius norm, the AED is not a differentiable function in all the points of its domain, so we use it as a metric rather than a loss. It also quantifies the similarity between two matrices with the precision with which $\hat{A}$ and $\hat{B}$ can be asserted to be the same matrix (and thus indicates the number of significant digits in the converged matrices $\hat{A}$, $\hat{B}$, etc). Using the AEDs, we identify subsets of identical branches (i.e., subsets $\{ \hat{W}_{\alpha}\}$ such that $d (\hat{W}_{\alpha}, \hat{W}_{\beta}) = 0$). Then, we retain a single representative from each subset, and discard the rest.


\subsection{Verifying the results of the ESM}\label{sec:verifying}



We can further confirm the symmetry group discovery by leveraging two fundamental group properties \cite{arfken2011mathematical}: (1) a symmetry group is closed under multiplication, and (2) the inverse of every group element must also belong to the group. Since these properties were not explicitly enforced during ESM optimization, they can be used to validate whether the identified set of matrices forms a group representation. For this purpose, we employ a group metric similar to the one proposed in \cite{calvo2024finding}.

The group metric is defined as
\begin{equation}
  \label{eq:d_group} d_{\mathrm{group}} = d_{\mathrm{closed}} + d_{\mathrm{inverse}},
\end{equation}
where the ``closed metric'' is given by
\begin{equation}
  \label{eq:d_clo} d_{\mathrm{closed}} = \frac{1}{M^2} \sum_{\alpha = 1}^M \sum_{\beta = 1}^M \min_{\gamma} d \left( \hat{W}_{\alpha} \hat{W}_{\beta} , \hat{W}_{\gamma} \right),
\end{equation}
and the ``inverse metric'' is defined as
\begin{equation}
  \label{eq:d_inv} d_{\mathrm{inverse}} = \frac{1}{M} \sum_{\alpha = 1}^M \min_{\gamma} d \left( \hat{W}_{\alpha}^{-1} , \hat{W}_{\gamma} \right).
\end{equation}
The group metric is a non-negative function of the branch matrices $\{ \hat{W}_{\alpha} \}$ that equals zero if and only if the set of matrices forms a group.


\section{Illustration of the ESM training method}\label{sec:ESM_proc}


In this section, we illustrate the proposed ESM training procedure using a simple example. We choose a dynamical system governed by
\begin{equation}\label{eq:cubic_decay}
  \frac{dx}{dt} = -x^3,
\end{equation}
i.e., $y(x) = -x^3$. The solutions are analytically known ($x(t)=\pm 1/\sqrt{2t + C}$, being $C$ an arbitrary constant), and decay asymptotically to $x=0$ as $t \rightarrow \infty$. This system is symmetric under sign inversion: if $x(t)$ is a solution, so is $-x(t)$. Thus, the symmetry group has $K=2$ elements: $D_1 = 1$ (identity) and $D_2 = -1$, that is $x' = D_{\alpha}x$ leaves Eq.~\ref{eq:cubic_decay} invariant. Note that, in this simple case, the transformations are scalars.


To generate the dataset, we simulate $\ntr = 10$ trajectories $\{ x^{(r)} (t_s) \}$ with time discretization of $\Delta t = 10^{-3}$ at $\nst = 100$ steps, using the Runge-Kutta method. The initial condition of each trajectory is randomly sampled from a uniform distribution within the interval $x(t=0) \in \left[-2.5,2.5 \right]$ (the initial condition determines the constant $C$ for each trajectory). We illustrate the method in two cases: (1) when $y_i$ is known analytically and (2) when only the trajectories are known and $y_i$ is estimated from them. We randomly shuffle the 1000 training pairs and use 750 samples for the oracle $y_{\mathrm{NN}}$ (see \ref{sec:train_det} for further details) and $N = 250$ for training the ESM.


\begin{figure*}
  \centering
  \includegraphics[width=\textwidth]{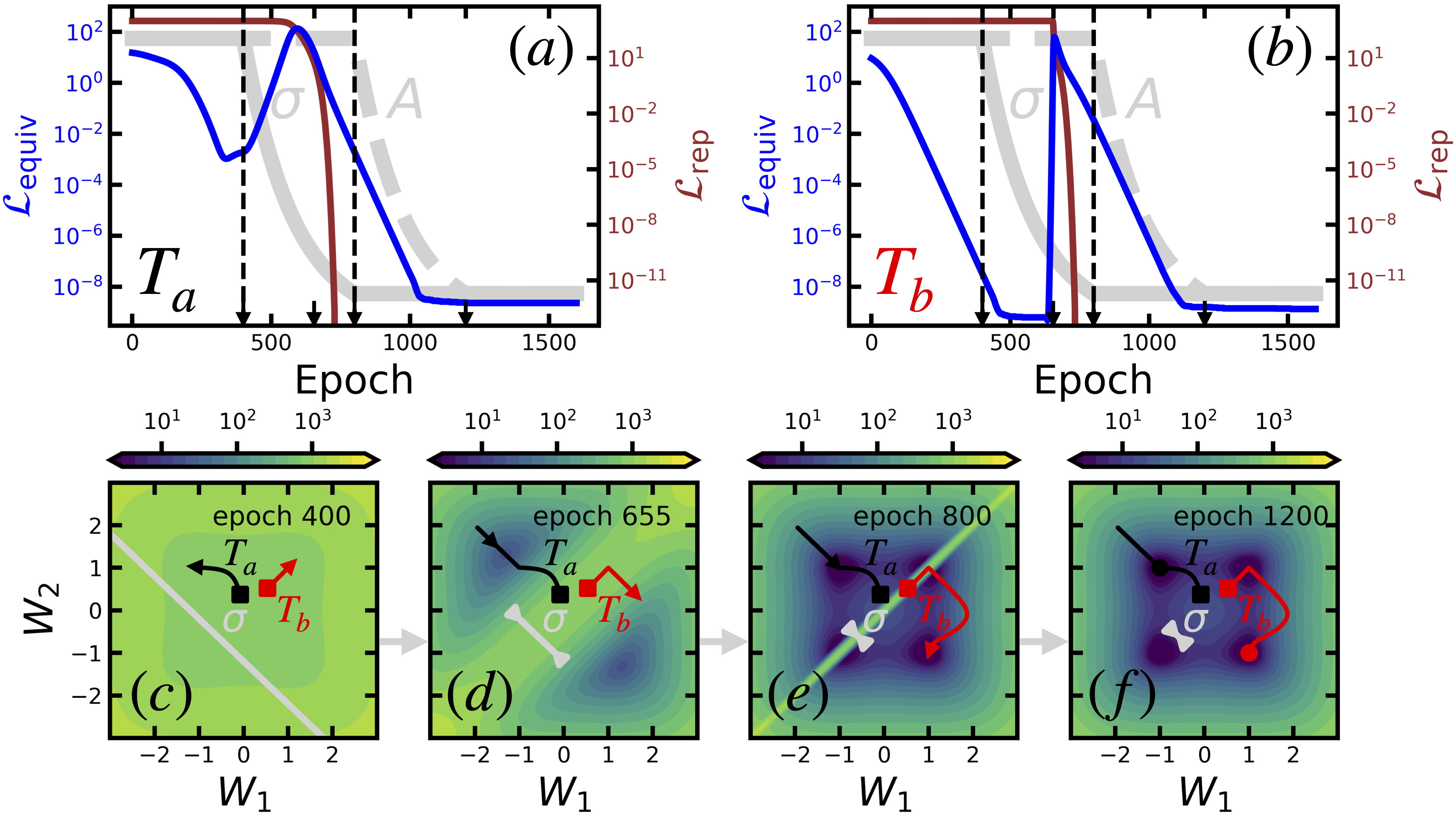}
  \caption{Illustration of ESM algorithm for the example $\dot{x} = - x^3$. The symmetry group has $K = 2$ elements: $D_1 = 1$ (identity) and $D_2 = -1$ in this case. We present two different trainings ($T_a$ and $T_b$) with different random initialization of the trainable parameters of the model, the branch weights $(W_1 , W_2)$. (a-b) Equivariance (blue) and repetition (brown) loss curves as function of epochs for $T_a$ and $T_b$ respectively. Gray curves represent the $\sigma$ (solid) and $A$ (dashed) values as function of the epochs (both normalized to the same arbitrary scale). The dashed vertical lines indicate the epochs at which each variation interval starts: first, $\sigma$ (from $\sigma_0 = 10^4$ to $\delta\sigma = 10^{-2}$) and second, $A$ (from $A_0 = 10^3$ to $\delta A = 10^{-8}$). (c-f) Colour maps of the total loss in the $(W_1 , W_2)$ plane at epochs 400, 655, 800 and 1200 respectively [these epochs are marked with black arrows in both horizontal axes of panels (a) and (b)]. The color bars represent the value of $\mathcal{L} = \lequivt + \lrep$. In all panels, trajectories in the "space of weights" [coordinates ($W_1$,$W_2$)] in $T_a$ and $T_b$ are plotted in black and red respectively. At each panel, these trajectories are plotted from the initial training location (squares) to the current state at a given epoch (triangles). In panel (f) (final epoch), the final results are represented by circles. The gray segments represent the Gaussian width in the definition of $\lrep$, corresponding to the current value of the hyperparameter $\sigma$. }
  \label{fig:FIG_2}
\end{figure*}


The process begins with fixed initial values $\sigma = \sigma_0$ and $A = A_0$, and the gradual elimination of $\lrep$ is executed in two stages. In the first stage, $\sigma$ is progressively reduced to $\sigma = \delta\sigma \approx 0$, which vanishes the influence of $\lrep$ for branch pairs $\hat{W}_\alpha$ and $\hat{W}_\beta$ located in different global minima (we do not set $\delta\sigma = 0$ because that would introduce a discontinuity in $\lrep$). In the second stage, $A$ is decreased to $A = \delta A \approx 0$, eliminating any remaining contribution of $\lrep$, even for redundant branches within the neighborhood of the same minimum, leaving only $\lequivt$ to drive the optimization (setting $\delta A \neq 0$ is convenient for reasons that become evident later in this section). As we will show below, these hyperparameter reduction processes allow knowing whether the number of branches has exceeded the order $K$ or not. If $M>K$, the procedure ends with the branch-removal processes explained in subsection \ref{sec:local_redundant}, culminating with an ESM having $M=K$ branches. In addition to this training, we validate the $K$-order symmetry group finding by calculating the group metric exposed in subsection \ref{sec:verifying}.


Fig. \ref{fig:FIG_2} shows the training procedure of a 2-branch ESM ($M = K$) with the analytical derivative function. The loss function $\mathcal{L}$ depends on the branch weights $(W_1, W_2)$, and $\lequivt$ has four global minima corresponding to the combinations of those in $\lequivs$: $(1, -1)$ and $(-1, 1)$, which capture the full symmetry group, and $(1, 1)$ and $(-1, -1)$, which leave one matrix undetected. We perform two trainings ($T_a$ and $T_b$, panels (a) and (b) in Fig.~\ref{fig:FIG_2}) with different weight initialization in order to demonstrate the method's effectiveness under various conditions. In both, hyperparameters $\sigma$ and $A$ are decreased exponentially (see \ref{sec:train_det}): the width runs from $\sigma_0 = 10^4$ to $\delta\sigma = 10^{-2}$ (between epochs 400 and 800), and the amplitude runs from $A_0 = 10^3$ to $\delta A = 10^{-8}$ (between epochs 800 and 1200).


Fig.~\ref{fig:FIG_2}(a) shows the evolution of the equivariance and repetition losses during training for setup $T_a$, while Fig.~\ref{fig:FIG_2}(b) does the same for setup $T_b$. Besides, Figs. \ref{fig:FIG_2}(c-f) visualize the total loss $\mathcal{L}$ landscapes [color maps of the total loss in the $(W_1, W_2)$ plane] at four representative epochs. The branch weight evolutions for both ESM trainings are traced from the initial configuration to the current one at each key epoch.

Initially, and up until epoch 400, when $\sigma$ begins to diminish, the large $\sigma$ value makes $\lrep$ nearly constant, leaving the optimization solely influenced by $\lequivt$. As a result, both trainings converge to isolated $\lequivt$ minima [see Fig.~\ref{fig:FIG_2}(c)]. However, $T_a$ converges to $(-1, 1)$, which is a desired situation where the two symmetry matrices are found, while $T_b$ converges to $(1, 1)$, which is an undesired situation where one symmetry matrix is repeated and the other is not found. During the process of $\sigma$ reduction from epochs 400 to 800, as shown in Fig.~\ref{fig:FIG_2}(a) and (b), the repetition loss begins to reshape the loss landscape, as it can be observed at epoch 655 in Fig.~\ref{fig:FIG_2}(d), driving the branches to reduce $\lrep$ while increasing $\lequivt$ in order to minimize the total loss. In $T_a$, the predicted branch weights undergo only slight perturbations as the $\mathcal{L}$ minimum $(-1,1)$ shifts, reflecting the weak repulsion between separated branch matrices due to high $\sigma$. In contrast, $\lrep$ transforms $(1,1)$ into a maximum, causing the branch weights in $T_b$ to move away from it. Since the branch matrices are repeated, they experience a strong repulsion, which pushes the second matrix toward the alternative $\lequivs$ minimum, $-1$.

By the end of the $\sigma$ reduction [epoch 800, see Fig. \ref{fig:FIG_2}(e)], the branch weights in both trainings are fully separated into distinct symmetry minima. At this stage, non-zero values of $\lrep$ appear only close to the diagonal $W_1 = W_2$. Note that local minima of $\mathcal{L}$ appear around the diagonal at $(1,1)$ and $(-1,-1)$. However, the dynamic $\sigma$-process has prevented the training to end up in these minima. Finally, as $A$ decreases from epochs 800 to 1200, $\lrep$ vanishes across the entire landscape, seen both in the evolution of the loss [Fig.~\ref{fig:FIG_2}(a) and (b)] and the landscape at epoch 1200 (panel (f), same figure). Since $M = K$, this variation has no effect on the optimization of the branch matrices because they are already located in minima of distinct symmetries.

Importantly, this experiment with $M=K$ demonstrates that, regardless of the weight initialization, the ESM training procedure always identifies the entire symmetry group, without detecting the same symmetry twice.

\begin{figure}
  \centering
  \includegraphics[width=\textwidth]{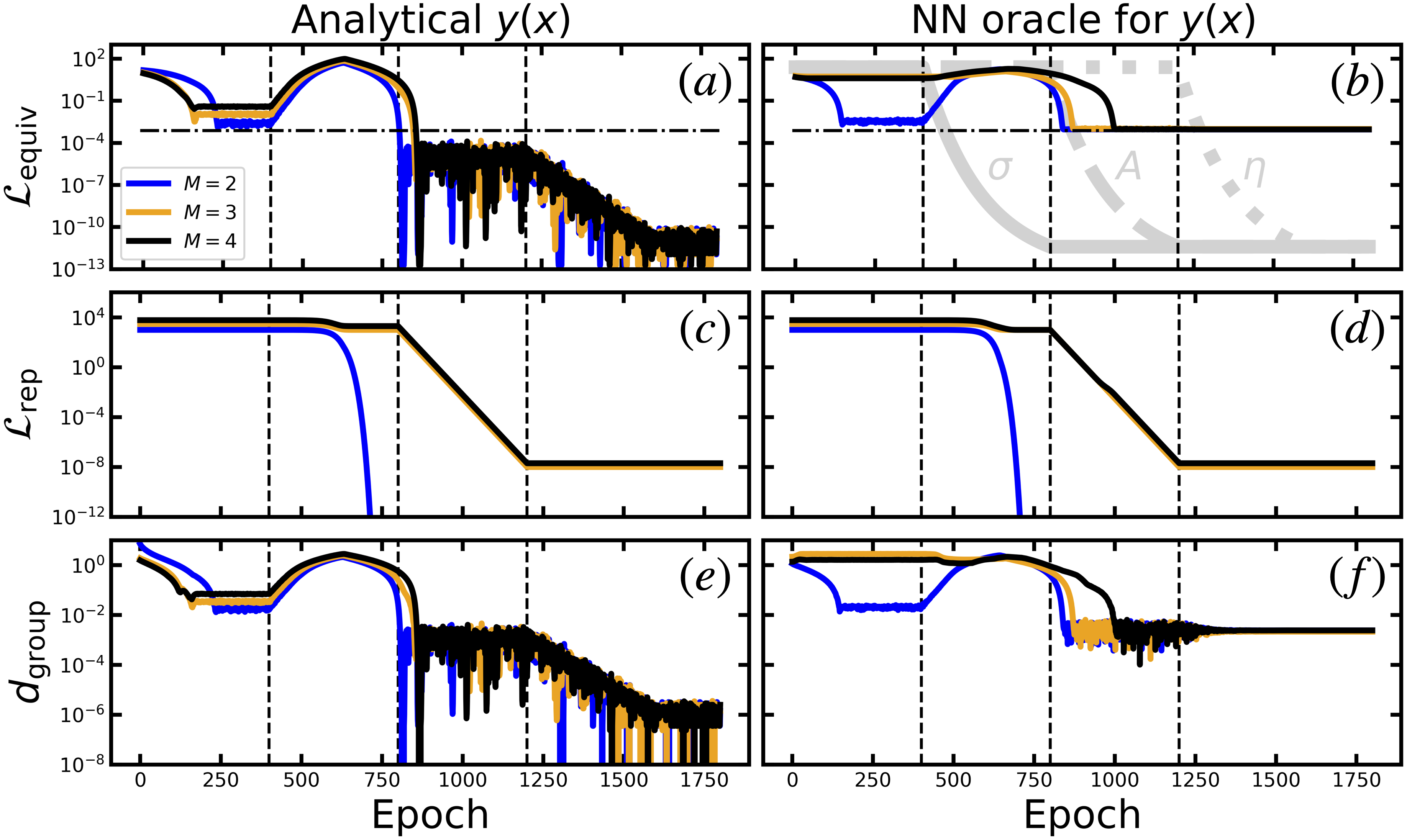}
  \caption{ESM algorithm for the illustrative system $\dot{x} = -x^3$ for different number of branches: $M = 2$ (blue), $M = 3$ (orange) and $M = 4$ (black). In the left column, the analytical function $\vec{y}(\vec{x})$ is used in the ESM. In the right column, a pretrained oracle $\vec{y}_{\mathrm{NN}} (\vec{x})$ is used instead. The gray curves in (b) provide a representation of the dependence of the hyperparameters $\sigma$ (solid), $A$ (dashed) and $\eta$ (dotted) as function of epochs (they are all normalized to the same arbitrary scale). In all panels, the dashed vertical lines mark the epoch at which a hyperparameter starts to decrease. (a-b) Equivariance loss as function of the epochs. The horizontal dashed-dotted line indicates the value $\lid$ given by the NN oracle's MSE. (c-d) Repetition loss as function of the epochs. (e-f) Group metric (Eq~\ref{eq:d_group}) as function of the epochs.}
  \label{fig:FIG_3}
\end{figure}



The variation of $A$ becomes crucial when $M>K$. We demonstrate this by considering ESMs with different number of branches: $M = 2, 3$ and $4$ (see Fig. \ref{fig:FIG_3}). For each $M$, we perform two trainings, one using the analytical derivative function and another using a pretrained oracle [$\mathcal{Y}(x) = y_{\mathrm{NN}} (x)$]. The variation stages for $\sigma$ and $A$ are identical to those in Fig. \ref{fig:FIG_2}. In addition to the steps previously followed, the learning rate $\eta$ is exponentially reduced from $\eta_0 = 10^{-3}$ to $\delta\eta = 10^{-6}$ between epochs 1200 and 1600 in order to mitigate gradient descent oscillations around the achieved minimum and, thus, improve convergence precision.


Fig.~\ref{fig:FIG_3}(a) shows the equivariance loss curves for each number of branches using $y(x)$, while Fig.~\ref{fig:FIG_3}(b) renders the same quantities when using $y_{\mathrm{NN}} (x)$ as oracle. Similarly, Figs. \ref{fig:FIG_3}(c) and \ref{fig:FIG_3}(d) show the corresponding repetition loss curves.


When $M = 2$, the behavior is consistent with Fig. \ref{fig:FIG_2}. At the end of the $\sigma$ variation process, $\lequivt$ is minimized and $\lrep$ vanishes, indicating that the ESM has identified $M$ distinct symmetries. However, when $M = 3$ or $M = 4$ ($M > K$), $\lrep$ does not vanish. In this case, since the number of branches has exceeded the number of unique global minima, some matrices inevitably fall into the same basin of attraction. These overlapping matrices interact via $\lrep$, even at low $\sigma$, introducing perturbations that prevent the total minimization of $\lequivt$ and, thus, an accurate prediction of the ground-truth symmetries. This difference in behavior between $M = K$ and $M > K$ cases provides a criterion for identifying the order of the symmetry group $K$ when it is unknown.


The importance of reducing parameter $A$ becomes evident at this stage. For $M = 3$, the solution converges to $\{ W_{\alpha} \} = \{ 1, -1, -1 \}$, where one ground-truth symmetry is repeated, resulting in a final $\lrep = \delta A$. For $M = 4$, the solution becomes $\{ W_{\alpha} \} = \{ 1, -1, 1, -1 \}$, with both ground-truth symmetries repeated, yielding $\lrep = 2\delta A$ [see Figs. \ref{fig:FIG_3}(c)-(d)]. Although matrix repetition occurs, it is no longer an issue as all global minima, and thus all symmetry elements $D_{\alpha}$, are identified.


Notably, for all $M$, the equivariance losses reach their expected bound ($0$ and $\lid$), confirming that the identified transformations are valid symmetries. A clear distinction can be observed in Figs. \ref{fig:FIG_3}(a) and \ref{fig:FIG_3}(b) between training with the analytical derivative function and the NN oracle. For ESMs trained with $\mathcal{Y}(x) = y(x)$, $\lequivt$ achieves extremely low values (around $10^{-12}$, which is much lower than the tolerance $\varepsilon_{\mathrm{equiv}}$) during the $\eta$-decreasing stage, demonstrating that the equivariance loss can be arbitrarily minimized. In contrast, for ESMs trained with $\mathcal{Y}(x) = y_{\mathrm{NN}}(x)$, $\lequivt$ consistently converges to the $\lid$ value given by the NN oracle's prediction error and cannot be further minimized, even as $\eta$ is reduced to enhance convergence precision.


For each value of $M$, we track the group metric (Eq.~\ref{eq:d_group}) as a function of epochs, as shown in Figs. \ref{fig:FIG_3}(e-f). In all cases, $d_{\mathrm{group}}$ decreases alongside $\lequivt$: the better the matrices satisfy the symmetry conditions, the better the set satisfies the group properties. This is particularly evident for $M = K = 2$, where $W_1 = 1$ and $W_2 = -1$ naturally fulfill the group criteria. When $M > K$, the group metric converges to a similarly low value, confirming that all symmetries are captured and redundant matrices maintain group closure and inverse properties.


Up to this point, we have demonstrated that the ESM reliably identifies all $K$ symmetry group elements, regardless of the weight initialization or the number of branches $M$ (provided that $M \geq K$). In simpler systems (e.g., Eq. \ref{eq:cubic_decay}), no additional processing is needed. The optimization is not affected by local minima in $\lequivs$, and redundant matrices in cases where $M > K$ can be easily identified by inspection. However, for more complex systems with higher dimensionality, solutions may occur at local minima, and also, recognizing duplicated matrices becomes less straightforward. For that cases, it would be necessary to apply the branch-removal processes explained in subsection \ref{sec:local_redundant}.


\section{Application of the ESM to complex dynamical systems}\label{sec:app_compl}

In this section, we showcase several examples of the ESM method applied to dynamical systems with higher complexity than the simple example considered in section \ref{sec:ESM_proc}, including some exhibiting chaotic behavior. As illustrations, we use Thomas symmetric attractor, Lorenz system, and a system of coupled Duffing oscillators.

The data generation is similarly conducted for all systems. The states $\{ \vec{x}^{(r)} (t_s) \}$ are obtained by generating $\ntr = 1000$ trajectories using the Runge-Kutta method, each with $\nst = 100$ time steps ($\Delta t = 10^{-3}$). Initial conditions are randomly sampled from a uniform distribution within the region of $\mathbb{R}^n$ that satisfies $-2.5 < x_k (t=0) < 2.5$. Derivatives $\{ \vec{y}^{(r)} (t_s) \}$ are obtained both analytically and numerically as in the previous section. After random shuffling, we use 95000 data pairs to train and validate the oracles $\ynn$ (see the \ref{sec:train_det} for details), and $N = 5000$ pairs to train the ESM.


We used the same hyperparameter settings in all these systems as well (see the \ref{sec:train_det} for a detailed summary of used hyperparameters). In each case, we apply three hyperparameter-tuning stages: the width $\sigma$ is reduced from $\sigma_0 = 10^{-1}$ to $\delta\sigma = 10^{-4}$ (epochs 600–1200), the amplitude $A$ from $A_0 = 10^{6}$ to $\delta A = 10^{-8}$ (epochs 1200–1800), and the learning rate $\eta$ from $\eta_0 = 10^{-3}$ to $\delta \eta = 10^{-6}$ (epochs 1800–2400). Branch-removal processes are conducted at epoch 2600 for local minima and at epoch 2800 for redundant branches.


\subsection{Thomas' symmetric attractor}\label{sec:thomas}

\begin{figure*}
  \centering
  \includegraphics[width=\textwidth]{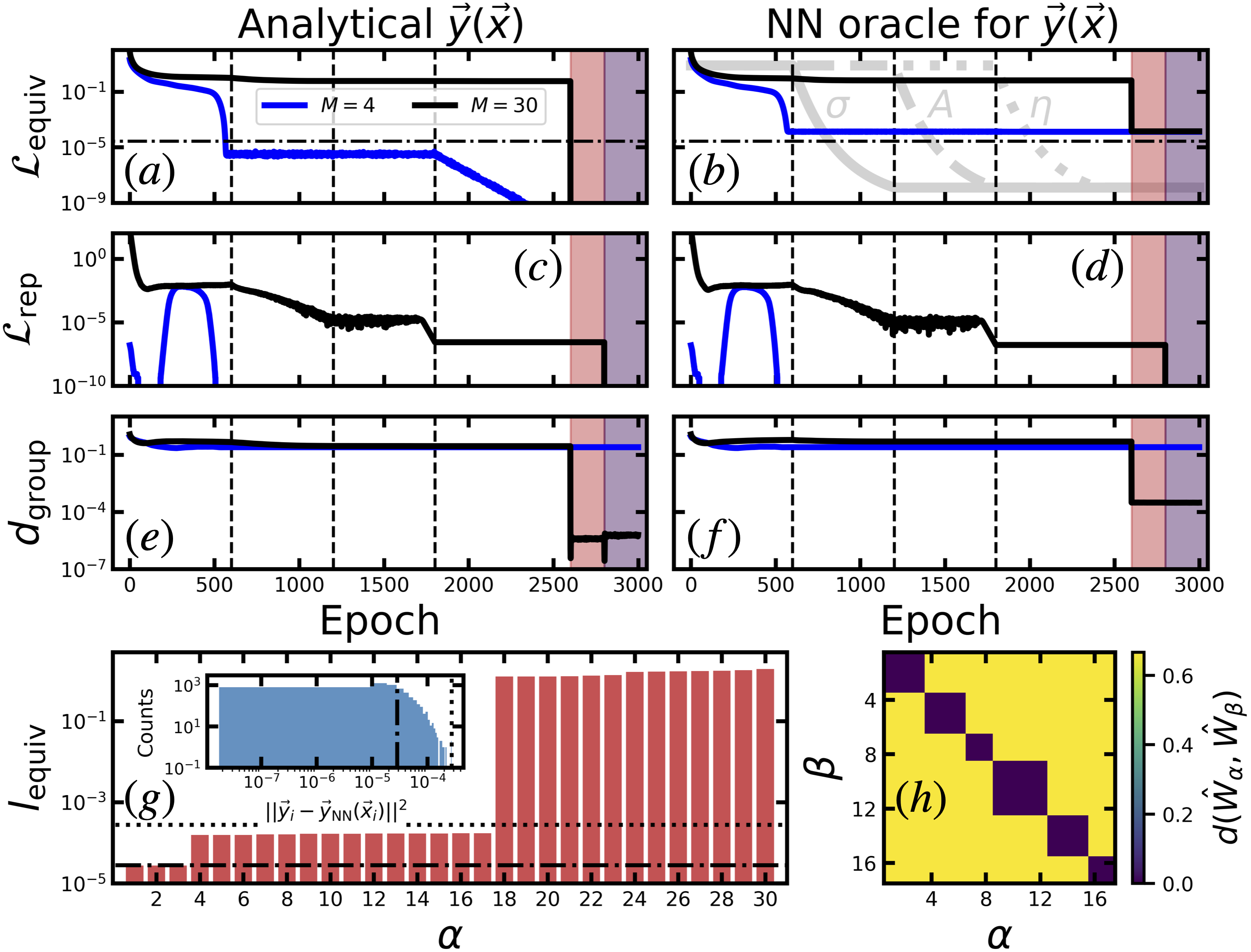}
  \caption{ESM algorithm for the Thomas' cyclically symmetric attractor. (a-f) Optimization of the matrices $\{ \hat{W}_{\alpha} \}$, using both the analytical function $\vec{y}(\vec{x})$ and oracle $\vec{y}_{\mathrm{NN}}(\vec{x})$, for $M = 4$ (blue) and $M = 30$ (black). Three quantities are shown as function of the epochs: equivariance loss (a-b), repetition loss (c-d) and group metric (e-f). The gray curves in (b) indicate the variation of $\sigma$, $A$ and $\eta$, and the vertical dashed lines in all panels represent the starting of each hyperparameter decrease. The dashed-dotted lines in (a-b) indicate the bound $\lid$ given by the oracle's MSE. In all panels, the red and purple areas indicate the regimes at which some branches have been removed by each of the two final-training processes. (g-h) Branch-removal processes for $M = 30$ using $\vec{y}_{\mathrm{NN}} (\vec{x})$. (g) Removal of matrices in local minima. The plot shows each single-matrix equivariance loss for all the matrices before the removal. The dotted-dashed line represents $\lid$, and the dotted line represents the tolerance $\varepsilon_{\mathrm{equiv}}$  given by the maximum oracle squared error. The inset displays the histogram of squared errors for all ESM training samples, with the dashed-dotted line representing $\lid$ the dotted line representing $\varepsilon_{\mathrm{equiv}}$. (h) Removal of redundant matrices. The plot shows the AED between all possible matrix pairs before the removal.}
  \label{fig:FIG_4}
\end{figure*}

Let us start with the Thomas' symmetric attractor. It is a three-dimensional system which was proposed by René Thomas \cite{thomas1999deterministic} and is described by the following equations:
\\

\begin{equation}\label{eq:thomas}
  \eqalign{ \frac{d x_1}{dt} =  \sin{x_2} - b x_1 \cr
  \frac{d x_2}{dt} =  \sin{x_3} - b x_2 \cr
  \frac{d x_3}{dt} =  \sin{x_1} - b x_3}
\end{equation}
Depending on the parameter $b$, this dynamical system exhibits various regimes, including chaotic behavior \cite{sprott2007labyrinth} (we used $b = 0.208186$ to be in the chaotic regime). These equations are symmetric under cyclical permutations of the state variables $(x_1, x_2, x_3)$ and under sign inversion (i.e., if $\vec{x}(t)$ is a solution, so is $-\vec{x}(t)$). Together, these symmetries form a group of order $K = 6$.

The results of the ESM trainings are shown in Fig. \ref{fig:FIG_4}. We consider two ESMs: one with $M = 4$ branches (to analyze the ESM performance when selecting an insufficiently large $M$ hyperparameter) and another with $M = 30$ branches (to significantly exceed $K$). Figs. \ref{fig:FIG_4}(a-f) show the evolution of the $\lequivt$, $\lrep$, and $d_{\mathrm{group}}$ for both values of $M$. Branch removal processes for the ESM with $M=30$ are indicated by the red and purple shaded regions. As in the previous section, we compare trainings that use either the analytical function $\vec{y}(\vec{x})$ from Eq. \ref{eq:thomas} [Figs. \ref{fig:FIG_4}(a,c,e)] or the NN oracle $\ynn$ [Figs. \ref{fig:FIG_4}(b,d,f)].


Let us first focus on the ESM with $M = 4$ branches. Immediately before the variation of $\sigma$, we observe that the equivariance loss is significantly minimized while the repetition loss vanishes. This indicates that the ESM has identified four distinct symmetry transformations (so final-training branch removals are not necessary in this case). However, the group metric does not decrease in parallel with $\lequivt$, suggesting that the set $\{ \hat{W}_{\alpha}\}$ does not form a group, and thus some symmetry transformations of the dynamical system have not been found.

In general, $d_{\mathrm{group}}$ is an appropriate measure to reveal whether the number of branches $M$ is sufficient to determine all symmetry transformations. Indeed, if $M$ is lower than the true symmetry group order $K$, it is likely that the found matrices do not form a group. The possible exception is when the set of matrices form a subgroup, which would cause the group metric to decrease significantly even when $M < K$. To identify whether a subgroup has been found, the ESM should be considered with one extra branch. If the converged $\lrep$ still vanishes, the previous set was a subgroup and some symmetry matrices were missing.


Now, consider the $M = 30$ case. As expected, $\lrep$ does not vanish after decreasing the Gaussian width $\sigma$, indicating that $K$ has been exceeded and some matrices are repeated. Unlike previous experiments, the minimization of $\lequivt$ does not converge to its lower bound after the Gaussian amplitude variation because some matrices $\hat{W}_{\alpha}$ converge to local minima that do not correspond to exact symmetries. This can be seen in Fig.~\ref{fig:FIG_4}(g) for the $\vec{\mathcal{Y}}(\vec{x}) = \ynn (\vec{x})$ case. The panel shows the single-branch equivariance losses for each $\alpha$ before the first removal, ordered in ascending $\lequivs$ values. We show with a horizontal dashed-dotted line the approximate bound of the function, $\lid \simeq 3 \cdot 10^{-5}$. With the dotted line, we show the tolerance $\varepsilon_{\mathrm{equiv}} \simeq 3 \cdot 10^{-4}$ that we use to remove local minima branches, whose value is given by the maximum oracle's error. As an inset, we represent the histogram of those oracle's errors $|| \vec{y}_i - \ynn (\vec{x}_i) ||^2$. Similarly, the vertical dashed-dotted and dotted lines represent $\lid$ and $\varepsilon_{\mathrm{equiv}}$ respectively. We observe that 13 branch matrices converge to local minima with $\lequivs (\hat{W}_{\alpha}) \sim 1$ (several orders of magnitude higher than $\lid$). The remaining 17 branches converge to global minima according to our criterion. It can be seen that  three out of these 17 global minima matrices have $\lequivs$ values extremely close to $\lid$ because they converge to the identity transformation. The others, associated to non-trivial symmetry transformations $\hat{D}_{\alpha}$, have slightly higher $\lequivs$ values, but still within the oracle's error distribution. 


The 13 local minima matrices are eliminated during the first removal process. In Figs. \ref{fig:FIG_4}(a) and \ref{fig:FIG_4}(b), we observe that $\lequivt$ drops abruptly after this removal. In the case of $\vec{\mathcal{Y}} (\vec{x}) = \vec{y} (\vec{x})$, $\lequivt$ completely vanishes. On the other hand, for $\vec{\mathcal{Y}} (\vec{x}) = \ynn (\vec{x})$, the equivariance loss decreases to a value similar to that achieved with the 4-branch ESM, which is consistent with the $\lequivs$ values in Fig.~\ref{fig:FIG_4}(g).

Finally, Fig.~\ref{fig:FIG_4}(h) displays the AEDs $d(\hat{W}_{\alpha}, \hat{W}_{\beta})$ before the second removal in the $\ynn$ case. In this figure, branch indices $\alpha, \beta$ are arranged in such a way that the first corresponds to the lowest $\lequivs$ value, and subsequent indices are sorted by proximity (the lowest AED to the previous branch). We can observe that the 17 remaining branch matrices contained 6 unique symmetry transformations. Removing redundant matrices makes $\lrep$ drop to zero [see Figs.~\ref{fig:FIG_4}(c) and (d)], confirming that no matrices are repeated. Importantly, $d_{\mathrm{group}}$ remains unchanged [see Figs.~\ref{fig:FIG_4}(e) and (f)], verifying that the removed matrices were redundant and the remaining set still forms a valid symmetry group. At the end of the process, the ESM accurately predicts the full finite symmetry group with $M = K = 6$ branch matrices, whose values are shown in \ref{sec:mat_res}.

\subsection{Lorenz system}\label{sec:lorenz}

\begin{figure*}
  \centering
  \includegraphics[width=\textwidth]{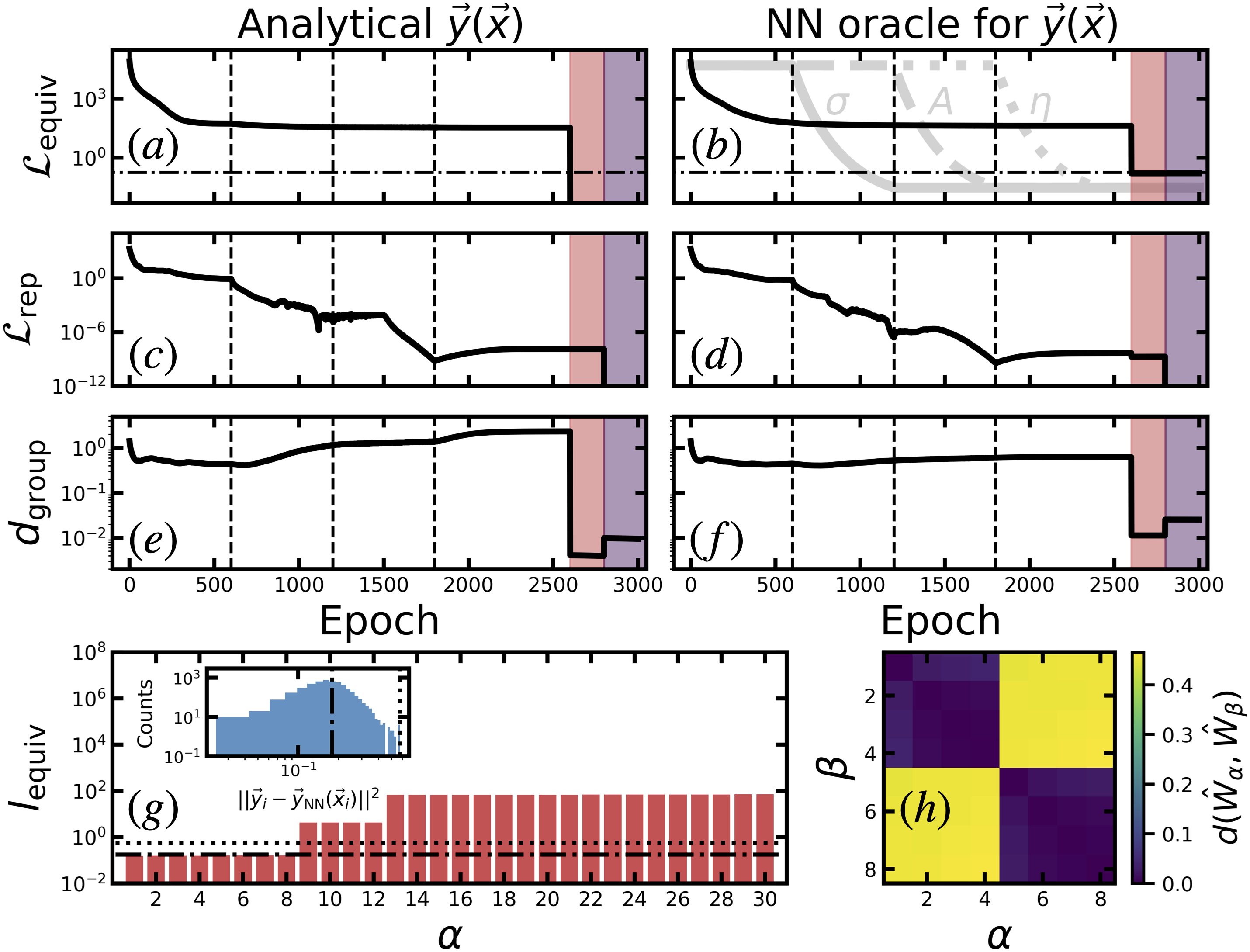}
  \caption{ESM algorithm with $M = 30$ branches for the Lorenz system. Three quantities are shown as function of the epochs: equivariance loss (a-b), repetiton loss (c-d) and group metric (e-f). Panels (g-h) illustrate the final-training processes using $\vec{y}_{\mathrm{NN}} (\vec{x})$. The remaining details are similar to those in Fig.~\ref{fig:FIG_4}.}
  \label{fig:FIG_5}
\end{figure*}

We consider now the Lorenz system, which is a three-dimensional system described by the following equations:
\begin{equation}\label{eq:lorenz}
  \eqalign{ \frac{d x_1}{dt} =  a (x_2 - x_1) \cr
  \frac{d x_2}{dt} =  x_1 (b - x_3) - x_2 \cr
  \frac{d x_3}{dt} =  x_1 x_2 - c x_3}
\end{equation}
being $a$, $b$ and $c$ are fixed parameters (we used $a = 10$, $b = 28$ and $c = 8/3$). This is a paradigmatic example of chaotic dynamics and was originally developed as a simplified model of atmospheric convection \cite{lorenz1963deterministic}.

The equations are symmetric under the sign inversion of the first two state variables, i.e., if $\vec{x} (t) = \left[ x_1 (t) , x_2 (t), x_3 (t) \right]^T$ is a solution, so is $\vec{x}' (t) = \left[ -x_1 (t) , -x_2 (t), x_3 (t) \right]^T$. Thus, the symmetry group has order $K = 2$.

The results are presented in Fig.~\ref{fig:FIG_5} with the same structure as in Fig.~\ref{fig:FIG_4}. This time, we explored only the $M>K$ case: we trained an ESM with $M = 30$ branches, both using the analytical function $\vec{y}(\vec{x})$ and the NN oracle $\ynn$. Figs. \ref{fig:FIG_5}(a-f) show the evolution of $\lequivt$, $\lrep$ and $d_{\mathrm{group}}$, and Figs. \ref{fig:FIG_5}(g-h) specifies the branch-removal processes for the case $\vec{\mathcal{Y}}(\vec{x}) = \ynn (\vec{x})$. We can observe that the oracle's error distribution sets a bound of the function $\lequivs$ given by $\lid \simeq 0.2$, with a tolerance of $\varepsilon_{\mathrm{equiv}} \simeq 0.6$.

We observe that the ESM training presents a similar behaviour to the previous case: after decreasing $\sigma$, $\lrep$ does not vanish because matrix repetition occurs inevitably, as $M>K$. Besides, $\lequivt$ does not vanish either because 22 matrices $\hat{W}_{\alpha}$ converge to local minima. We observe in Fig.~\ref{fig:FIG_5}(g) (NN oracle case) that the values of $\lequivs(\hat{W}_{\alpha})$ for those $\hat{W}_{\alpha}$ in local minima are of the order of $10^{1}$ and $10^{2}$, much higher than $\lid$. This matrices are pruned in the first branch-removal process, producing $\lequivt$ to drop to its lower bound ($\lequivt \simeq 0$ and $\lequivt \simeq \lid$). We see in Fig.~\ref{fig:FIG_5}(h) that the 8 global minima branches contain only 2 unique symmetries. The second branch-removal process prunes 6 matrices, leaving the ESM with $M = K = 2$ (see \ref{sec:mat_res} for obtained values).


\subsection{Coupled Duffing oscillators}\label{sec:duffing}

\begin{figure*}
  \centering
  \includegraphics[width=\textwidth]{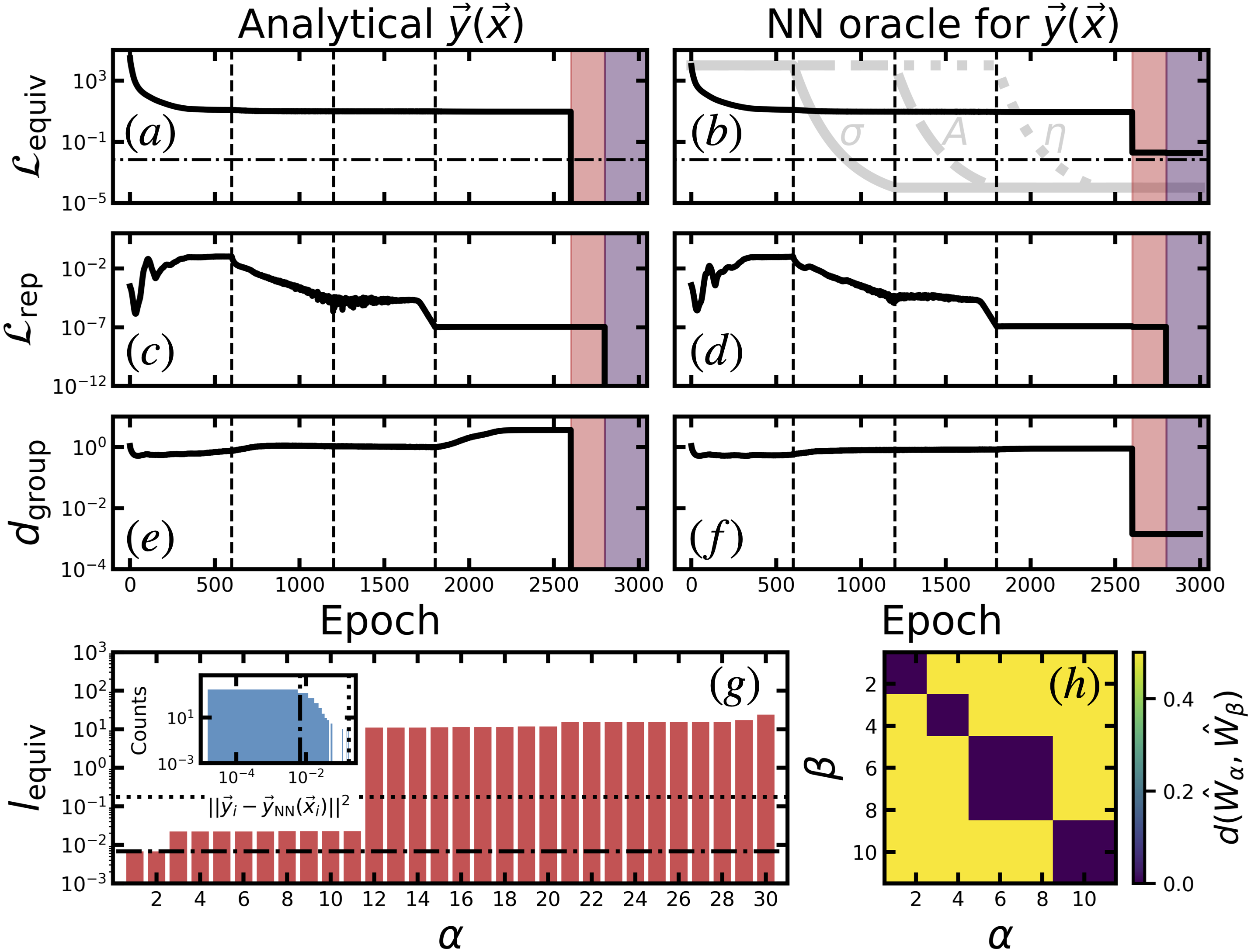}
  \caption{ESM algorithm with $M = 30$ branches for the coupled Duffing oscillators. Three quantities are shown as function of the epochs: equivariance loss (a-b), repetiton loss (c-d) and group metric (e-f). Panels (g-h) illustrate the final-training processes using $\vec{y}_{\mathrm{NN}} (\vec{x})$. The remaining details are similar to those in Fig.~\ref{fig:FIG_4}.}
  \label{fig:FIG_6}
\end{figure*}

The last example is a four-dimensional dynamical system that consists of two coupled Duffing oscillators, whose equations of motion are
\begin{equation}\label{eq:duffing}
  \eqalign{ \frac{d x_1}{dt} =  x_2 \cr
  \frac{d x_2}{dt} =  -\delta x_2 - \alpha x_1 - \beta x_1^3 + \gamma (x_1 - x_3) \cr
  \frac{d x_3}{dt} =  x_4 \cr
  \frac{d x_4}{dt} =  -\delta x_4 - \alpha x_3 - \beta x_3^3 + \gamma ( x_3 - x_1 ) }
\end{equation}
where $x_1$ and $x_2$ denote the position and velocity of the first oscillator respectively, whereas $x_3$ and $x_4$ those of the second. The fixed parameters are: $\delta$ (damping), $\alpha$, $\beta$ (linear and non-linear restoring forces), and $\gamma$ (coupling constant). For the numerical experiment, we used $\delta = 0.3$, $\alpha = -1$, $\beta = 1$ and $\gamma = 5$. The system exhibits symmetry under sign inversion [if $\vec{x} (t)$ is a solution, so is $-\vec{x} (t)$] and permutation of oscillators (if $\vec{x} (t) = \left[ x_1 (t) , x_2 (t), x_3 (t) , x_4 (t) \right]^T$ is a solution, so is $\vec{x}' (t) = \left[ x_3 (t) , x_4 (t), x_1 (t) , x_2 (t) \right]^T$). Together, these form a symmetry group of order $K = 4$.

The results are shown in Fig.~\ref{fig:FIG_6} with the same structure as in previous cases. We trained an ESM with $M = 30$ branches, also in the $\vec{\mathcal{Y}} (\vec{x}) = \vec{y} (\vec{x})$ and $\vec{\mathcal{Y}} (\vec{x}) = \ynn (\vec{x})$ cases. As we see, the ESM behavior is consistent with those presented in previous subsections. This time, in the NN oracle case [see Fig~\ref{fig:FIG_6}(g)], the $\lequivs$ bound is given by $\lid \simeq 7 \cdot 10^{-3}$, with a tolerance of $\varepsilon_{\mathrm{equiv}} \simeq 0.2$. We see that 19 branches converge to local minima (with values $\lequivs(\hat{W}_{\alpha}) \sim 10^1$), and 11 branches converge to global minima with $\lequivs$ values that lie within the oracle's error distribution. To conclude, we observe in Fig.{\ref{fig:FIG_6}}(h) that, at the end of the protocole, 4 distinct symmetries where detected, coinciding with the order $K=4$ (see \ref{sec:mat_res} for the obtained values).


\section{Conclusions}\label{sec:conc}

In this work, we introduced the ESM, a ML framework designed to discover the full finite group of linear symmetries in arbitrary nonlinear dynamical systems. The method focuses on identifying equivariant transformations through a training process regularized by the repetition loss, which ensures the complete identification of all symmetries within the group. Unlike iterative approaches, the ESM unveils all these elements in a single training cycle, with the group metric validating the results.

Notably, the ESM effectively works in both theory-informed and purely data-driven settings. While it can leverage analytical equations when available, it requires only trajectory data to operate, with no prior knowledge of the system’s governing equations. This is achieved by pretraining an oracle approximator to estimate the system's dynamics. In our experiment, we employed a simple feed-forward NN as the oracle, but the method is adaptable and can integrate alternative ML methods.

\section*{Acknowledgments}

We acknowledge projects PID2023-148359NB-C21 and CEX2023-001286-S (financed by MICIU/AEI/10.13039/501100011033) and the Government of Aragon through Project Q-MAD.

\section*{Code availability}

The code for this paper can be found in the following Github repository: \url{https://github.com/pablocalvo7/Equivariance_Seeker_Model.git}.

\appendix


\section{Training details}\label{sec:train_det}

This Appendix details the training procedures for the experiments in this paper, including ESM training and NN oracle pre-training. All calculations were performed using TensorFlow \cite{tensorflow2015-whitepaper} and Keras \cite{chollet2015keras} libraries.

For all ESM experiments, hyperparameter variation followed the formula:
\begin{equation}
  \label{cases}
  h(\mathrm{epoch})=\cases{h_0&for $\mathrm{epoch} < \mathrm{epoch}_i$\\
  h_0 e^{ -\omega (\mathrm{epoch} - \mathrm{epoch}_i )}&for $\mathrm{epoch}_i \leq \mathrm{epoch} \leq \mathrm{epoch}_f$\\
  \delta h&for $\mathrm{epoch} > \mathrm{epoch}_f$\\}
\end{equation} 
with
\begin{equation}
  \omega = \frac{1}{\mathrm{epoch}_f - \mathrm{epoch}_i} \log{ \frac{h_0}{\delta h} },
\end{equation}
where $h$ represents the tuned hyperparameter (either $\sigma$, $A$ or $\eta$), $h_0$ and $\delta h$ denote the initial and final hyperparameter values, and $ \left[ \mathrm{epoch}_i , \mathrm{epoch}_f \right]$ specifies the variation interval. These values, as well as further hyperparameters and training information, are summarized in Table \ref{tab:hyp_sum} for each experiment in this paper. Note that the experiments illustrated in Figs. \ref{fig:FIG_4}, \ref{fig:FIG_5} and \ref{fig:FIG_6} share identical training setting.

\begin{table}
  \caption{Summary of ESM training hyperparameters from each experiment presented in this article. This involves four different systems: the illustrative one-dimensional system (equation \ref{eq:cubic_decay}), Thomas system (equation \ref{eq:thomas}), Lorenz system (equation \ref{eq:lorenz}) and the coupled Duffing oscillators (equation \ref{eq:duffing}).}
  \label{tab:hyp_sum}
  \begin{tabular*}{\textwidth}{@{}l*{15}{@{\extracolsep{0pt plus
  12pt}}l}}
  \br
  Hyperparameter&Figure 2&Figure 3&Figure 4, 5 and 6\\
  \mr
  $( \sigma_0 , \delta\sigma )$&$( 10^4 , 10^{-2} )$&$( 10^4 , 10^{-2} )$&$( 10^{-1} , 10^{-4} )$\\
  $( A_0 , \delta A )$&$( 10^{3} , 10^{-8} )$&$( 10^{3} , 10^{-8} )$&$( 10^{6} , 10^{-8} )$\\
  $( \eta_0 , \delta \eta )$&$10^{-4}$ (no variation)&$( 10^{-3} , 10^{-6} )$&$( 10^{-3} , 10^{-6} )$\\
  Variation interval $\sigma$&400-800&400-800&600-1200\\
  Variation interval $A$&800-1200&800-1200&1200-1800\\
  Variation interval $\eta$&-&1200-1600&1800-2400\\
  Epoch 1st removal&-&-&2600\\
  Epoch 2nd removal&-&-&2800\\
  Total epochs&1200&1800&3000\\
  N (ESM data)&250&250&5000\\
  Optimizer&SGD&RMSprop&RMSprop\\
  Mini-batch size&64&64&1024\\

  \br
  \end{tabular*}
\end{table}

We trained four different feedforward NN oracles $\ynn$, one for each dynamical system considered in this paper. As mentioned in the main text, these NNs approximate the derivative field, so $\ynn (\vec{x}) \simeq \vec{y}(\vec{x})$. For that, we train them using a subset of randomly chosen pairs from the data set $\{ \vec{x}^{(r)} (t_s), \vec{y}^{(r)}(t_s) \}$, with $\vec{y}^{(r)}(t_s)$ computed numerically as explained in section \ref{sec:math_def}. All oracles were trained using the MSE loss and the Adam optimizer.

\begin{figure*}
  \centering
  \includegraphics[width=\textwidth]{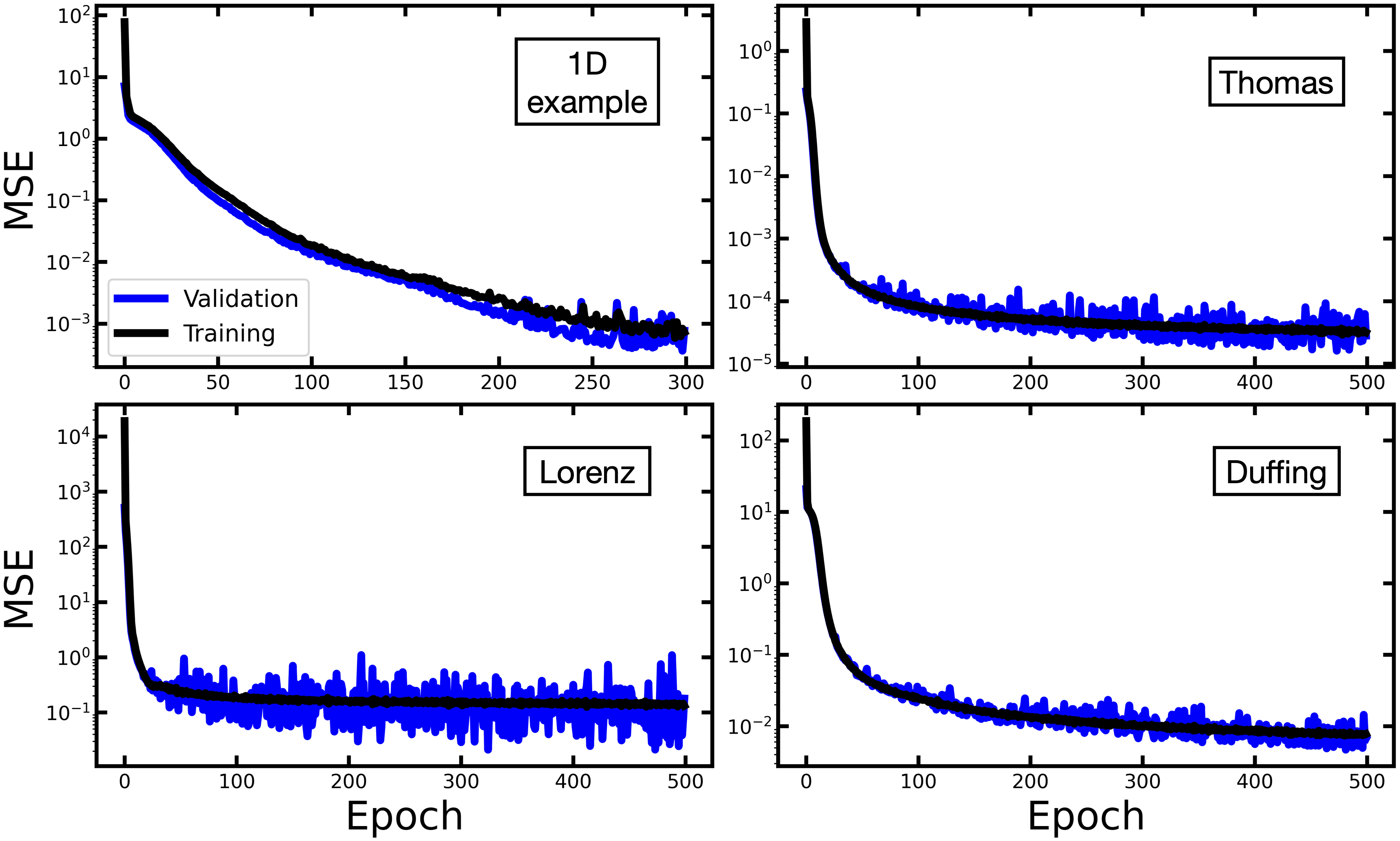}
  \caption{Training of four NN oracles $\ynn$ prior to ESM training. Each panel correspond to a dynamical system and shows the oracle's MSE loss as function of epochs for training (black curve) and validation (blue curve) data sets.}
  \label{fig:FIG_AP}
\end{figure*}

We now summarize each oracle training hyperparameters and settings:
\begin{itemize}
  \item 1D illustrative system (Eq. \ref{eq:cubic_decay}): 500 samples for training and 250 for validation. The architecture is composed by the input layer ($n=1$ neuron), 2 hidden layers (100 neurons) with ReLU activation function and the output layer ($n=1$ neuron) with linear activation function. The NN was trained during 300 epochs using a learning rate of $10^{-4}$ and a mini-batch size of 32.
  \item Thomas, Lorenz and Duffing systems (Eqs. \ref{eq:thomas}, \ref{eq:lorenz} and \ref{eq:duffing}): 90000 samples were used for training and 5000 for validation. The architecture is composed by the input layer comprising either 3 (Thomas and Lorenz) or 4 (Duffing) neurons, 3 hidden layers (500 neurons each) with ReLU activation function and the output layer composed of 3 or 4 neurons, with linear activation function. The NN was trained during 500 epochs using a learning rate of $10^{-5}$ and a mini-batch size of 256.
\end{itemize}
Oracle training curves are shown in Fig. \ref{fig:FIG_AP}. Once trained, the oracle weights are fixed and integrated into the $\vec{\mathcal{Y}}$ block of the ESMs.

\section{Branch matrix results}\label{sec:mat_res}

In this Appendix, we present the matrix values of the symmetry groups discovered using our method in Section \ref{sec:app_compl}, that is, for the Thomas attractor, Lorenz system and the Duffing oscillators. Specifically, we provide in Table \ref{tab:mat_res} the numerical results obtained from the ESMs with $M = 30$ branches in the case of $\vec{\mathcal{Y}}(x) = \ynn(\vec{x})$, truncated to five decimal places.

In general, the appropiate decimal precision for rounding the obtained matrix values should be chosen such that the group metric $d_{\mathrm{group}}$ is minimized. Notably, in all cases presented in Table \ref{tab:mat_res}, rounding the matrix values to the nearest integer yields the exact symmetry matrices $\{ \hat{D}_{\alpha} \}$, ensuring that $d_{\mathrm{group}}$ becomes exactly zero.

\begin{table}
  \caption{Obtained matrix values $\{ \hat{W}_{\alpha} \}$ using ESMs with $M = 30$ branches and NN oracles for the systems presented in section \ref{sec:app_compl}: Thomas system (equation \ref{eq:thomas}), Lorenz system (equation \ref{eq:lorenz}) and the coupled Duffing oscillators (equation \ref{eq:duffing}).}
  \scriptsize
  \label{tab:mat_res}
  \begin{tabular*}{\textwidth}{@{}l*{15}{@{\extracolsep{0pt plus
  12pt}}l}}
  \br
  Thomas&Lorenz&Duffing\\
  \mr
  $\left(\begin{array}{ccc}
    1.00034 & -0.00003 & -0.00002 \\
    0.00006 &  1.00001 & -0.00008 \\
   -0.00014 & -0.00003 &  0.99983
  \end{array}\right)$&$\left(\begin{array}{ccc}
    0.95049 &  0.06961 & -0.00038 \\
    0.02487 &  0.97252 & -0.00023 \\
    0.00095 & -0.00043 &  0.99007
  \end{array}\right)$&$\left(\begin{array}{cccc}
    -0.00036 &  0.00051 & -1.00051 & -0.00172 \\
     0.00068 &  0.00228 &  0.00024 & -0.99846 \\
    -0.99871 & -0.00416 &  0.00109 &  0.00327 \\
    -0.00013 & -0.99827 &  0.00049 &  0.00125
   \end{array}\right)$\\

  \\
  $\left(\begin{array}{ccc}
    -0.00018 & -1.00007 & -0.00016 \\
    -0.00053 &  0.00014 & -1.00011 \\
    -1.00073 &  0.00007 &  0.00017
  \end{array}\right)$&$\left(\begin{array}{ccc}
    -0.96805 & -0.05061 &  0.00096 \\
    -0.01825 & -0.98413 &  0.00012 \\
     0.00070 & -0.00052 &  0.99607
   \end{array}\right)$&$\left(\begin{array}{cccc}
    \phantom{-}0.99990 &  0.00061 & -0.00032 & -0.00216 \\
    \phantom{-}0.00040 &  1.00076 &  0.00009 &  0.00005 \\
    \phantom{-}0.00014 & -0.00221 &  1.00057 & -0.00100 \\
    \phantom{-}0.00015 &  0.00073 &  0.00019 &  0.99999
  \end{array}\right)$\\

  \\
  $\left(\begin{array}{ccc}
    -0.00019 &  0.00020 & -1.00025 \\
    -1.00009 & -0.00008 & -0.00006 \\
    -0.00033 & -1.00014 & -0.00005
  \end{array}\right)$&\phantom{$\left(\begin{array}{ccc}
    xxxxxxxx & xxxxxxxx & xxxxxxxx \\
    xxxxxxxx &  xxxxxxxx & xxxxxxxx \\
   xxxxxxxx & xxxxxxxx &  xxxxxxxx
  \end{array}\right)$}&$\left(\begin{array}{cccc}
    -0.00013 & -0.00049 &  1.00048 & -0.00110 \\
     0.00037 & -0.00034 &  0.00004 &  1.00050 \\
     1.00062 & -0.00053 & -0.00142 & -0.00137 \\
     0.00020 &  1.00082 &  0.00009 & -0.00081
   \end{array}\right)$\\
  
   \\
  $\left(\begin{array}{ccc}
    -1.00048 &  0.00011 &  0.00037 \\
     0.00004 & -1.00032 & -0.00023 \\
    -0.00037 &  0.00016 & -1.00021
  \end{array}\right)$&\phantom{$\left(\begin{array}{ccc}
    xxxxxxxx & xxxxxxxx & xxxxxxxx \\
    xxxxxxxx &  xxxxxxxx & xxxxxxxx \\
   xxxxxxxx & xxxxxxxx &  xxxxxxxx
  \end{array}\right)$}&$\left(\begin{array}{cccc}
    -0.99945 &  0.00026 &  0.00017 & -0.00120 \\
     0.00045 & -0.99925 &  0.00057 &  0.00217 \\
     0.00021 & -0.00318 & -0.99978 &  0.00154 \\
    -0.00022 &  0.00124 &  0.00039 & -0.99861
   \end{array}\right)$\\
  
  \\
  $\left(\begin{array}{ccc}
    0.00009 & -0.00017 &  1.00008 \\
    1.00069 & -0.00003 & -0.00001 \\
   -0.00001 &  1.00044 &  0.00005
  \end{array}\right)$&\phantom{$\left(\begin{array}{ccc}
    xxxxxxxx & xxxxxxxx & xxxxxxxx \\
    xxxxxxxx &  xxxxxxxx & xxxxxxxx \\
   xxxxxxxx & xxxxxxxx &  xxxxxxxx
  \end{array}\right)$}&\phantom{$\left(\begin{array}{cccc}
    xxxxxxxx &  xxxxxxxx &  xxxxxxxx & xxxxxxxx \\
    xxxxxxxx & xxxxxxxx &  xxxxxxxx &  xxxxxxxx \\
    xxxxxxxx & xxxxxxxx & xxxxxxxx &  xxxxxxxx \\
    xxxxxxxx &  xxxxxxxx &  xxxxxxxx & xxxxxxxx
   \end{array}\right)$}\\

  \\
  $\left(\begin{array}{ccc}
    -0.00011 &  1.00037 &  0.00056 \\
     0.00004 & -0.00020 &  1.00058 \\
     1.00014 & -0.00001 & -0.00013
    \end{array}\right)$&\phantom{$\left(\begin{array}{ccc}
      xxxxxxxx & xxxxxxxx & xxxxxxxx \\
      xxxxxxxx &  xxxxxxxx & xxxxxxxx \\
     xxxxxxxx & xxxxxxxx &  xxxxxxxx
    \end{array}\right)$}&\phantom{$\left(\begin{array}{cccc}
      xxxxxxxx &  xxxxxxxx &  xxxxxxxx & xxxxxxxx \\
      xxxxxxxx & xxxxxxxx &  xxxxxxxx &  xxxxxxxx \\
      xxxxxxxx & xxxxxxxx & xxxxxxxx &  xxxxxxxx \\
      xxxxxxxx &  xxxxxxxx &  xxxxxxxx & xxxxxxxx
     \end{array}\right)$}\\

  \br
  \end{tabular*}
\end{table}

\clearpage

\section*{References}

\begin{thebibliography}{10}

  \bibitem{gross1996therole}
  David~J. Gross.
  \newblock The role of symmetry in fundamental physics.
  \newblock {\em Proceedings of the National Academy of Sciences},
    93(25):14256--14259, 1996.
  
  \bibitem{noether1918invariante}
  E.~Noether.
  \newblock Invariante variationsprobleme.
  \newblock {\em Nachrichten von der Gesellschaft der Wissenschaften zu
    Göttingen, Mathematisch-Physikalische Klasse}, 1918:235--257, 1918.
  
  \bibitem{wang2023scientific}
  Hanchen Wang, Tianfan Fu, Yuanqi Du, Wenhao Gao, Kexin Huang, Ziming Liu, Payal
    Chandak, Shengchao Liu, Peter Van~Katwyk, Andreea Deac, Anima Anandkumar,
    Karianne Bergen, Carla~P. Gomes, Shirley Ho, Pushmeet Kohli, Joan Lasenby,
    Jure Leskovec, Tie-Yan Liu, Arjun Manrai, Debora Marks, Bharath Ramsundar,
    Le~Song, Jimeng Sun, Jian Tang, Petar Veličković, Max Welling, Linfeng
    Zhang, Connor~W. Coley, Yoshua Bengio, and Marinka Zitnik.
  \newblock Scientific discovery in the age of artificial intelligence.
  \newblock {\em Nature}, 620(7972):47–60, 2023.
  
  \bibitem{yu2024learning}
  Rose Yu and Rui Wang.
  \newblock Learning dynamical systems from data: An introduction to
    physics-guided deep learning.
  \newblock {\em Proceedings of the National Academy of Sciences},
    121(27):e2311808121, 2024.
  
  \bibitem{brunton2016discovering}
  Steven~L. Brunton, Joshua~L. Proctor, and J.~Nathan Kutz.
  \newblock Discovering governing equations from data by sparse identification of
    nonlinear dynamical systems.
  \newblock {\em Proceedings of the National Academy of Sciences},
    113(15):3932--3937, 2016.
  
  \bibitem{rudy2017datadriven}
  Samuel~H. Rudy, Steven~L. Brunton, Joshua~L. Proctor, and J.~Nathan Kutz.
  \newblock Data-driven discovery of partial differential equations.
  \newblock {\em Science Advances}, 3(4):e1602614, 2017.
  
  \bibitem{liu2021machine}
  Ziming Liu and Max Tegmark.
  \newblock Machine learning conservation laws from trajectories.
  \newblock {\em Phys. Rev. Lett.}, 126:180604, May 2021.
  
  \bibitem{liu2022machine}
  Ziming Liu, Varun Madhavan, and Max Tegmark.
  \newblock Machine learning conservation laws from differential equations.
  \newblock {\em Phys. Rev. E}, 106:045307, Oct 2022.
  
  \bibitem{liu2024interpretable}
  Ziming Liu, Patrick~Obin Sturm, Saketh Bharadwaj, Sam~J. Silva, and Max
    Tegmark.
  \newblock Interpretable conservation laws as sparse invariants.
  \newblock {\em Phys. Rev. E}, 109:L023301, Feb 2024.
  
  \bibitem{ha2021discovering}
  Seungwoong Ha and Hawoong Jeong.
  \newblock Discovering invariants via machine learning.
  \newblock {\em Phys. Rev. Res.}, 3:L042035, Dec 2021.
  
  \bibitem{wetzel2020discovering}
  Sebastian~J. Wetzel, Roger~G. Melko, Joseph Scott, Maysum Panju, and Vijay
    Ganesh.
  \newblock Discovering symmetry invariants and conserved quantities by
    interpreting siamese neural networks.
  \newblock {\em Phys. Rev. Res.}, 2:033499, Sep 2020.
  
  \bibitem{lu2023discovering}
  Peter~Y Lu, Rumen Dangovski, and Marin Solja{\v{c}}i{\'c}.
  \newblock Discovering conservation laws using optimal transport and manifold
    learning.
  \newblock {\em Nature Communications}, 14(1):4744, 2023.
  
  \bibitem{iten2020discovering}
  Raban Iten, Tony Metger, Henrik Wilming, L\'{\i}dia del Rio, and Renato Renner.
  \newblock Discovering physical concepts with neural networks.
  \newblock {\em Phys. Rev. Lett.}, 124:010508, Jan 2020.
  
  \bibitem{greydanus2019hamiltonian}
  Samuel Greydanus, Misko Dzamba, and Jason Yosinski.
  \newblock Hamiltonian neural networks.
  \newblock {\em Advances in neural information processing systems}, 32, 2019.
  
  \bibitem{cranmer2020lagrangian}
  Miles Cranmer, Sam Greydanus, Stephan Hoyer, Peter Battaglia, David Spergel,
    and Shirley Ho.
  \newblock Lagrangian neural networks.
  \newblock {\em arXiv preprint arXiv:2003.04630}, 2020.
  
  \bibitem{liu2022machinelearninghidden}
  Ziming Liu and Max Tegmark.
  \newblock Machine learning hidden symmetries.
  \newblock {\em Phys. Rev. Lett.}, 128:180201, May 2022.
  
  \bibitem{bondesan2019learning}
  Roberto Bondesan and Austen Lamacraft.
  \newblock Learning symmetries of classical integrable systems.
  \newblock {\em arXiv preprint arXiv:1906.04645}, 2019.
  
  \bibitem{desai2022symmetry}
  Krish Desai, Benjamin Nachman, and Jesse Thaler.
  \newblock Symmetry discovery with deep learning.
  \newblock {\em Phys. Rev. D}, 105:096031, May 2022.
  
  \bibitem{yang2023generative}
  Jianke Yang, Robin Walters, Nima Dehmamy, and Rose Yu.
  \newblock Generative adversarial symmetry discovery.
  \newblock In Andreas Krause, Emma Brunskill, Kyunghyun Cho, Barbara Engelhardt,
    Sivan Sabato, and Jonathan Scarlett, editors, {\em Proceedings of the 40th
    International Conference on Machine Learning}, volume 202 of {\em Proceedings
    of Machine Learning Research}, pages 39488--39508. PMLR, 23--29 Jul 2023.
  
  \bibitem{barenboim2021symmetry}
  Gabriela Barenboim, Johannes Hirn, and Veronica Sanz.
  \newblock {Symmetry meets AI}.
  \newblock {\em SciPost Phys.}, 11:014, 2021.
  
  \bibitem{krippendorf2020detecting}
  Sven Krippendorf and Marc Syvaeri.
  \newblock Detecting symmetries with neural networks.
  \newblock {\em Machine Learning: Science and Technology}, 2(1):015010, dec
    2020.
  
  \bibitem{decelle2019learning}
  A.~Decelle, V.~Martin-Mayor, and B.~Seoane.
  \newblock Learning a local symmetry with neural networks.
  \newblock {\em Phys. Rev. E}, 100:050102, Nov 2019.
  
  \bibitem{gabel2023learning}
  Alex Gabel, Victoria Klein, Riccardo Valperga, Jeroen S.~W. Lamb, Kevin
    Webster, Rick Quax, and Efstratios Gavves.
  \newblock Learning lie group symmetry transformations with neural networks.
  \newblock In {\em Proceedings of 2nd Annual Workshop on Topology, Algebra, and
    Geometry in Machine Learning (TAG-ML)}, volume 221 of {\em Proceedings of
    Machine Learning Research}, pages 50--59. PMLR, 28 Jul 2023.
  
  \bibitem{gabel2024datadriven}
  Alex Gabel, Rick Quax, and Efstratios Gavves.
  \newblock Data-driven lie point symmetry detection for continuous dynamical
    systems.
  \newblock {\em Machine Learning: Science and Technology}, 5(1):015037, mar
    2024.
  
  \bibitem{ko2024learning}
  Gyeonghoon Ko, Hyunsu Kim, and Juho Lee.
  \newblock Learning infinitesimal generators of continuous symmetries from data.
  \newblock {\em arXiv preprint arXiv:2410.21853}, 2024.
  
  \bibitem{forestano2023deep}
  Roy~T Forestano, Konstantin~T Matchev, Katia Matcheva, Alexander Roman, Eyup~B
    Unlu, and Sarunas Verner.
  \newblock Deep learning symmetries and their lie groups, algebras, and
    subalgebras from first principles.
  \newblock {\em Machine Learning: Science and Technology}, 4(2):025027, jun
    2023.
  
  \bibitem{dehmamy2021automatic}
  Nima Dehmamy, Robin Walters, Yanchen Liu, Dashun Wang, and Rose Yu.
  \newblock Automatic symmetry discovery with lie algebra convolutional network.
  \newblock {\em Advances in Neural Information Processing Systems},
    34:2503--2515, 2021.
  
  \bibitem{hou2024machine}
  Wanda Hou, Molan Li, and Yi-Zhuang You.
  \newblock Machine learning symmetry discovery for classical mechanics.
  \newblock {\em arXiv preprint arXiv:2412.14632}, 2024.
  
  \bibitem{calvo2024finding}
  Pablo Calvo-Barl\'es, Sergio~G. Rodrigo, Eduardo S\'anchez-Burillo, and Luis
    Mart\'{\i}n-Moreno.
  \newblock Finding discrete symmetry groups via machine learning.
  \newblock {\em Phys. Rev. E}, 110:045304, Oct 2024.
  
  \bibitem{bronstein2017geometric}
  Michael~M. Bronstein, Joan Bruna, Yann LeCun, Arthur Szlam, and Pierre
    Vandergheynst.
  \newblock Geometric deep learning: Going beyond euclidean data.
  \newblock {\em IEEE Signal Processing Magazine}, 34(4):18--42, 2017.
  
  \bibitem{cohen2016group}
  Taco Cohen and Max Welling.
  \newblock Group equivariant convolutional networks.
  \newblock In Maria~Florina Balcan and Kilian~Q. Weinberger, editors, {\em
    Proceedings of The 33rd International Conference on Machine Learning},
    volume~48 of {\em Proceedings of Machine Learning Research}, pages
    2990--2999, New York, New York, USA, 20--22 Jun 2016. PMLR.
  
  \bibitem{satorras2021enequivariant}
  V{\i}ctor~Garcia Satorras, Emiel Hoogeboom, and Max Welling.
  \newblock E (n) equivariant graph neural networks.
  \newblock In {\em International conference on machine learning}, pages
    9323--9332. PMLR, 2021.
  
  \bibitem{cohen2019gauge}
  Taco Cohen, Maurice Weiler, Berkay Kicanaoglu, and Max Welling.
  \newblock Gauge equivariant convolutional networks and the icosahedral cnn.
  \newblock In {\em International conference on Machine learning}, pages
    1321--1330. PMLR, 2019.
  
  \bibitem{thomas2018tensor}
  Nathaniel Thomas, Tess Smidt, Steven Kearnes, Lusann Yang, Li~Li, Kai Kohlhoff,
    and Patrick Riley.
  \newblock Tensor field networks: Rotation-and translation-equivariant neural
    networks for 3d point clouds.
  \newblock {\em arXiv preprint arXiv:1802.08219}, 2018.
  
  \bibitem{salova2019koopman}
  Anastasiya Salova, Jeffrey Emenheiser, Adam Rupe, James~P. Crutchfield, and
    Raissa~M. D’Souza.
  \newblock Koopman operator and its approximations for systems with symmetries.
  \newblock {\em Chaos: An Interdisciplinary Journal of Nonlinear Science},
    29(9):093128, 09 2019.
  
  \bibitem{schmid2010dynamic}
  Peter~J. Schmid.
  \newblock Dynamic mode decomposition of numerical and experimental data.
  \newblock {\em Journal of Fluid Mechanics}, 656:5–28, 2010.
  
  \bibitem{chen2018neural}
  Ricky~TQ Chen, Yulia Rubanova, Jesse Bettencourt, and David~K Duvenaud.
  \newblock Neural ordinary differential equations.
  \newblock {\em Advances in neural information processing systems}, 31, 2018.
  
  \bibitem{brunton2019datadriven}
  Steven~L. Brunton and J.~Nathan Kutz.
  \newblock {\em Data-Driven Science and Engineering: Machine Learning, Dynamical
    Systems, and Control}.
  \newblock Cambridge University Press, 2019.
  
  \bibitem{arfken2011mathematical}
  George~B Arfken, Hans~J Weber, and Frank~E Harris.
  \newblock {\em Mathematical methods for physicists: a comprehensive guide}.
  \newblock Academic press, 2011.
  
  \bibitem{thomas1999deterministic}
  R.~Thomas.
  \newblock Deterministic chaos seen in terms of feedback circuits: Analysis,
    synthesis, "labyrinth chaos".
  \newblock {\em International Journal of Bifurcation and Chaos},
    09(10):1889--1905, 1999.
  
  \bibitem{sprott2007labyrinth}
  Julien~Clinton Sprott and Konstantinos~E Chlouverakis.
  \newblock Labyrinth chaos.
  \newblock {\em International Journal of Bifurcation and Chaos},
    17(06):2097--2108, 2007.
  
  \bibitem{lorenz1963deterministic}
  Edward~N Lorenz.
  \newblock Deterministic nonperiodic flow.
  \newblock {\em Journal of atmospheric sciences}, 20(2):130--141, 1963.
  
  \bibitem{tensorflow2015-whitepaper}
  Mart\'{i}n Abadi, Ashish Agarwal, Paul Barham, Eugene Brevdo, Zhifeng Chen,
    Craig Citro, Greg~S. Corrado, Andy Davis, Jeffrey Dean, Matthieu Devin,
    Sanjay Ghemawat, Ian Goodfellow, Andrew Harp, Geoffrey Irving, Michael Isard,
    Yangqing Jia, Rafal Jozefowicz, Lukasz Kaiser, Manjunath Kudlur, Josh
    Levenberg, Dandelion Man\'{e}, Rajat Monga, Sherry Moore, Derek Murray, Chris
    Olah, Mike Schuster, Jonathon Shlens, Benoit Steiner, Ilya Sutskever, Kunal
    Talwar, Paul Tucker, Vincent Vanhoucke, Vijay Vasudevan, Fernanda Vi\'{e}gas,
    Oriol Vinyals, Pete Warden, Martin Wattenberg, Martin Wicke, Yuan Yu, and
    Xiaoqiang Zheng.
  \newblock {TensorFlow}: Large-scale machine learning on heterogeneous systems,
    2015.
  \newblock Software available from tensorflow.org.
  
  \bibitem{chollet2015keras}
  Fran\c{c}ois Chollet et~al.
  \newblock Keras.
  \newblock \url{https://keras.io}, 2015.
  
  \end{thebibliography}

\end{document}